\documentclass[10pt,twocolumn,twoside]{IEEEtran} 
\usepackage{enumerate}
\usepackage{graphicx}
\usepackage{latexsym}
\usepackage{amsmath}
\usepackage{amssymb}
\usepackage{epsfig}
\usepackage{cite}
\usepackage{algorithmic}
\usepackage{overpic}
\usepackage{multirow}
\usepackage[table]{xcolor}
\usepackage{graphicx}
\usepackage{colortbl,array}
\usepackage{multirow,bigdelim}
\usepackage{graphicx}
\usepackage{subfigure}
\usepackage{booktabs}
\usepackage{booktabs,amsmath}
\usepackage{fancyhdr, graphicx, amsmath, amssymb}
\usepackage[linesnumbered,ruled]{algorithm2e}
\usepackage{arydshln}
\usepackage[english]{babel}
\usepackage{xcolor}

\newtheorem{rem}{Remark}
\newtheorem{asm}{Assumption}
\newtheorem{thm}{Theorem}
\newtheorem{cor}{Corollary}
\newtheorem{lem}{Lemma}
\newtheorem{definition}{Definition}
\newtheorem{prop}{Proposition}

\ifCLASSINFOpdf
\else
\fi

\begin{document}
\title{Strategic Topology Switching for Security--Part II: Detection \& Switching Topologies}

\author{Yanbing~Mao, Emrah~Akyol,
        and Ziang~Zhang
\IEEEcompsocitemizethanks{\IEEEcompsocthanksitem Y.~Mao, E.~Akyol and Z.~Zhang are with the Department of Electrical and Computer Engineering, Binghamton University--SUNY, Binghamton, NY,
13902 USA, (e-mail: \{ymao3, eakyol, zhangzia\}@binghamton.edu).
}
\thanks{}}

\maketitle

\begin{abstract}
This two-part paper considers strategic topology switching for security in the second-order multi-agent system. In Part II, we propose a strategy on switching topologies to detect zero-dynamics attack (ZDA), whose attack-starting time is allowed to be not the initial time. We first characterize the sufficient and necessary condition for detectability of ZDA, in terms of the network topologies to be switched to and the set of agents to be monitored. 
We then propose an attack detection algorithm based on the Luenberger observer, using the characterized detectability condition. Employing the strategy on switching times proposed in Part I~\cite{YA} and the strategy on switching topologies proposed here, a strategic topology-switching algorithm is derived. Its primary advantages are threefold: (i) in achieving consensus in the absence of attacks, the control protocol does not need velocity measurements and the algorithm has no constraint on the magnitudes of coupling weights; (ii) in tracking system in the absence of attacks, the Luenberger observer has no constraint on the magnitudes of observer gains and the number of monitored agents, i.e., only one monitored agent's output is sufficient; (iii) in detecting ZDA, the algorithm allows the defender to have no knowledge of the attack-starting time and the number of misbehaving agents (i.e., agents under attack). Simulations are provided to verify the effectiveness of the strategic topology-switching algorithm.
\end{abstract}

\begin{IEEEkeywords}
Multi-agent system, strategic topology switching, zero-dynamics attack, attack-starting time, attack detection.
\end{IEEEkeywords}
\IEEEpeerreviewmaketitle

\section{Introduction}
\IEEEPARstart{I}{n} Part-I paper~\cite{YA}, the proposed simplified control protocol under switching topology employs only relative positions of agents, which is different from the well-studied control protocols~\cite{YC10,A16x,MR16,A16,RE07,huang2016some,abdessameud2010consensus}. The main objective of this two-part paper is the strategic topology-switching algorithm for the second-order multi-agent system under attack. The algorithm is based on two strategies, one of which on switching times and the other on switching topologies. The strategy on switching times, as introduced in Part-I paper~\cite{YA}, enables the second-order multi-agent system in the absence of attacks to reach the second-order consensus. The strategy on switching topologies proposed in this Part-II paper enables the strategic topology-switching algorithm to detect stealthy attacks. 

Security concerns regarding the networked systems pose a formidable threat to their wide deployment, as highlighted by the recent incidents including distributed denial-of-service (DDOS) attack on Estonian web sites~\cite{nazario2009politically}, Maroochy water breach \cite{slay2007lessons} and  cyber attacks on smart grids \cite{cnn2}.  The  ``networked" aspect  exacerbates the difficulty of securing these systems since centralized measurement (sensing) and control are not feasible for such large-scale systems~\cite{F13}, and hence require the development of decentralized approaches, which are inherently prone to attacks.  Particularly,  a special class of "stealthy" attacks, namely the  ``zero-dynamics attack" (ZDA), poses a significant security challenge~\cite{MP15,hl18,MP18}. The main idea behind ZDA is to hide the attack signal  in the null-space of the state-space representation of the control system so that it cannot be detected by applying conventional detection methods on the observation signal (hence, the name ``stealthy").  The objective of such an attack can vary from manipulating the controller to accept false data that would yield the system towards a desired (e.g., unstable) state to maliciously altering system dynamics (topology attack~\cite{kim2013topology}) to affect system trajectory.

Recent research efforts have focused on variations of ZDA for systems with distinct properties. For stochastic cyber-physical systems, Park et al.~\cite{park2016} designed a robust ZDA, where the attack-detection signal is guaranteed to stay below a threshold over a finite horizon. In~\cite{kim2016zero}, Kim et al. proposed a discretized ZDA for the sampled-date control systems, where the attack-detection signal is constantly zero at the sampling times. Another interesting line of research pertains to developing defense strategies \cite{n11,n12,F13,weerakkody2017robust,chen2017protecting}.  For example, Jafarnejadsani et al.~\cite{jafarnejadsani2017dual,hl18} proposed a multi-rate $\mathcal{L}_{1}$ adaptive controller which can detect ZDA in the sampled-data control systems, by removing certain unstable zeros of discrete-time systems~\cite{MP15,MP18}. Back et al.~\cite{back2017enhancement}  utilized ``generalized hold" to render the impact of bounded ZDA.

While developing defense strategies for the ZDAs in multi-agent systems have recently gained interest~\cite{n11,n12,F13,weerakkody2017robust,chen2017protecting}  (see Table I for a brief summary), the space of solutions is yet to be thoroughly explored. The most prominent features of prior work are that the conditions of detectable attack have constrain the connectivity of network topology and the number of the misbehaving agents (referred to agents under attack)~\cite{n11,n12,F13,weerakkody2017robust}, and the corresponding developed defense strategy for attack detection works effectively only in situation where the number of misbehaving agents and the attack-starting time being the initial time are known to the defender~\cite{n11,n12,F13,T12,n15,chen2017protecting}. The main objective of this work is to remove such constraints and unrealistic assumptions by utilizing a new approach for attack detection: intentional topology switching.

Recent experiment of stealthy false-data injection attacks on networked control system~\cite{P12} showed the changes in the system dynamics could be utilized by defender to detect ZDA. To have changes in the system dynamics, Teixeira et al.~\cite{T12} proposed a  method of modifying output matrix through adding and removing observed measurements, or modifying input matrix through adding and removing actuators or perturbing the control input matrix. But the defense strategy requires the attack-starting times to be the initial time and known to defender. In other words, the defense strategy fails to work if the attack-starting time is designed to be not the initial time and the defender has no such knowledge, as is practically the case for most scenarios. In such realistic scenario, to detect ZDA, the system dynamics must have dynamic changes, i.e., some parameters of system dynamics changes infinitely over infinite time. However, before using the dynamic changes to detect ZDA in such realistic situation, the question that \emph{whether the dynamic changes in system dynamics can destroy system stability in the absence of attacks?} must be investigated. If the dynamic changes can destroy system stability in the absence of attacks, these changes could utilized by adversary/attacker~\cite{ZW16,KT13,ciftcioglu2017topology}.

In recent several years, actively/strategically topology switching has received significantly attention in the control theory, network science and graph theory literatures, see e.g., Amelkin and Singh~\cite{amelkin2017disabling} proposed edge recommendation to disable external influences of adversaries in social networks (consensus-seeking social dynamics), while the coupling weights are uncontrollable since they correspond to the users' interpersonal appraisals; Mao and Akyol~\cite{allerton18, mao2018synchronization} showed that strategic (time-dependent) topology switching is an effective method in detecting ZDA in the coupled harmonic oscillators; Ciftcioglu et al.~\cite{ciftcioglu2017topology} studied dynamic topology design in the adversary environment where the network designer continually and strategically change network topology to a denser state, while the adversary attempts to thwart the defense strategy simultaneously.

Moreover, driven by recent developments in mobile computing, wireless communication and sensing~\cite{hartenstein2008tutorial}, it is more feasible to set communication topology as a control variable~\cite{mazumder2011wireless}. These motivate us to consider the method of topology switching to induce changes in the dynamics of multi-agent systems to detect ZDA. The strategy on switching times proposed in Part-I paper~\cite{YA} answers the question: \emph{when the topology of network should switch such that the occurring dynamic changes in system dynamics do not undermine the agent's ability of reaching consensus in the absence of attacks}. Based on the work in Part-I paper~\cite{YA}, this Part-II paper focuses on the strategy on switching topologies that addresses the problem of \emph{switching to what topologies to detect ZDA}.

The contribution of this paper is fourfold, which can be summarized as follows.
\begin{itemize}
 \item A ZDA variation is first studied, whose attack-starting time can be not the initial time.
  \item We characterize the sufficient and necessary condition for detectability of the ZDA variation under strategic topology switching.
  \item We characterize the sufficient and necessary condition for Luenberger observer in tracking real multi-agent system in the absence of attacks, which has no constraint on the number of monitored agents.
  \item Based on the strategy on switching times and the strategy on switching topologies, through employing Luenberger observer, a strategic topology-switching algorithm for attack detection is proposed. The advantages of the algorithm are: i) in detecting ZDA, it allows the defender to have no knowledge of misbehaving agents and the attack-starting time; ii) in tracking real systems in the absence of attacks, it has no constraint on the magnitudes of observer gains and  the number of monitored agents; iii) in achieving the second-order consensus, it has no constraint on the magnitudes of coupling weights, while the control protocol does not need velocity measurements.
\end{itemize}

This paper is organized as follows. We present the preliminaries and the problem formulation in Sections II and III respectively. In Section IV, we characterize the condition for detectability of ZDA.  Based on this characterization, we develop an attack detection algorithm in Section V.  We provide numerical simulation results  in Section VI, and, in Section VII we discuss the future research directions. 
\begin{table}
\caption{Conditions on Detectable Attack}
\label{msft}
\centering
\scalebox{0.83}[0.83]{%
\begin{tabular}{  p{1.2cm}  |  p{6.0cm} |  p{2.0cm}}
\bottomrule
\cellcolor[gray]{0.9} \textbf{Reference} & \cellcolor[gray]{0.9} $ \cellcolor[gray]{0.9} \textbf{Conditions}$ &
\cellcolor[gray]{0.9} $\textbf{Dynamics}$\\  \hline
        \cite{n11}   &connectivity is not smaller than $2|\mathbb{K}|$ + 1 &Discrete Time\\ \hline
        \cite{n12} & $|\mathbb{K}|$ is smaller than connectivity &Discrete Time \\ \hline
        \cite{F13}  & size of input-output linking is smaller than $|\mathbb{K}|$  & Continuous Time\\ \hline
        \cite{weerakkody2017robust}  & the minimum vertex separator is larger than $|\mathbb{K}| + 1$  & Discrete Time\\ \hline

        \cite{chen2017protecting}  & single attack, i.e., $|\mathbb{K}|  = 1$  & Continuous Time\\
        \bottomrule
\end{tabular}
}
\end{table}

\section{Preliminaries}
\subsection{Notation}
We use $P < 0$ to denote a negative definite matrix $P$. $\mathbb{R}^{n}$ and $\mathbb{R}^{m \times n}$ denote the set of $\emph{n}$-dimensional real vectors and the set of $m \times n$-dimensional real matrices, respectively. Let $\mathbb{C}$ denote the set of complex number. $\mathbb{N}$ represents the set of the natural numbers and
$\mathbb{N}_{0}$ = $\mathbb{N}$ $\cup$ $\left\{ 0 \right\}$. Let $\mathbf{1}_{n \times n}$ and $\mathbf{0}_{n \times n}$ be the $n \times n$-dimensional identity matrix and zero matrix, respectively. $\mathbf{1}_{n} \in \mathbb{R}^{n}$ and $\mathbf{0}_{n} \in \mathbb{R}^{n}$ denote the vector with all ones and the vector with all zeros, respectively. The superscript `$\top$' stands for matrix transpose.

The interaction among $n$ agents is modeled by an undirected graph $\mathrm{G} = (\mathbb{V}, \mathbb{E})$, where
$\mathbb{V}$ = $\left\{ {1,2, \ldots, n} \right\}$ is the set of agents and
$\mathbb{E} \subset \mathbb{V} \times \mathbb{V}$ is the set of edges. The weighted adjacency matrix $\mathcal{A} = \left[ {{a_{ij}}} \right]$ $\in \mathbb{R}^{n \times n}$ is defined as $a_{ij} = a_{ji} > 0$ if $(i, j) \in \mathbb{E}$, and $a_{ij} = a_{ji} = 0$ otherwise. Assume that there are no self-loops, i.e., for any ${i} \in \mathbb{V}$, $a_{ii} = 0$.  The Laplacian matrix of an undirected graph $\mathrm{G}$ is defined as $\mathcal{L} = \left[ {{l_{ij}}} \right] \in {\mathbb{R}^{n \times n}}$, where ${l_{ii}} = \sum\limits_{j = 1}^n {{a_{ij}}}$ and ${l_{ij}} =  - {a_{ij}}$ for $i \neq j$.

Some important notations are highlighted as follows:
\begin{description}
  \item[$\emph{\emph{lcm}}(\cdot):$]  \hspace{0.4cm}operator of least common multiple among scalers;
  \item[$\ker \left( Q \right):$]     \hspace{0.4cm}set $\left\{ {y: Qy = {\mathbf{0}_n}}, Q \in \mathbb{R}^{n \times n} \right\}$;
  \item[$A^{-1}\mathbb{F}$]     \hspace{0.4cm}set $\left\{ {y: Ay \in \mathbb{F}}\right\}$;
  \item[$\left| \mathbb{V} \right|:$] \hspace{0.4cm}cardinality (i.e., size) of the set $\mathbb{V}$;
  \item[${\mathbb{V}} \backslash \mathbb{K}:$] \hspace{0.29cm} complement set of $\mathbb{K}$ with respect to $\mathbb{V}$;
  \item[$\lambda_{i}\left( M \right):$] \hspace{0.35cm} $i^{\emph{\emph{th}}}$ eigenvalue of matrix $M \in \mathbb{R}^{n \times n}$;
  \item[$\mathfrak{S}\left( r \right):$] \hspace{0.35cm} $r^{\emph{\emph{th}}}$ element of ordered set $\mathfrak{S}$;
  \item[$\mu_{P}(\cdot):$] \hspace{0.35cm} matrix measure of induced $P$-norm;
  \item[$\mathbb{Q}:$] \hspace{0.35cm} set of rational numbers.
\end{description}

\subsection{Attack Model}
As a class of stealthy attacks, ``zero-dynamics" attack is hard to detect, identify, and then mitigate from a control theory perspective~\cite{MP15,hl18,MP18}. Before reviewing its attack policy, let us first consider the following system:
\begin{subequations}
\begin{align}
\dot z\left( t \right) &= Az\left( t \right),\\
y\left( t \right) &= Cz\left( t \right),
\end{align}\label{eq:sl2}\end{subequations}
where $z(t) \in \mathbb{R}^{\bar{n}}$ and $y(t) \in \mathbb{R}^{\bar{m}}$ denote system state and monitored output, respectively; $A \in \mathbb{R}^{\bar{n} \times \bar{n}}$, $C \in \mathbb{R}^{\bar{m} \times \bar{n}}$. Its corresponding version under attack is described by
\begin{subequations}
\begin{align}
\dot{\tilde z}\left( t \right) &= A\tilde z\left( t \right) + Bg(t),\\
\tilde y\left( t \right) &= C\tilde z\left( t \right) + Dg(t),
\end{align}\label{eq:sl1}\end{subequations}
where ${g}(t) \in \mathbb{R}^{\bar{o}}$ is attack signal, $B \in \mathbb{R}^{\bar{n} \times \bar{o}}$ and $D \in \mathbb{R}^{\bar{m} \times \bar{o}}$.

The policy of ZDA with introduction of attack-starting time is presented in the following definition, which is different from the attack policies studied in~\cite{n11,n12,F13,T12,n15,chen2017protecting}, whose attack-starting times are the initial time.
\begin{definition} The attack signal
\begin{align}
{g}(t) = \left\{ \begin{array}{l}
\hspace{-0.2cm}{g}{e^{\eta \left( {t - \rho } \right)}},t \in \left[ {\rho,\infty } \right)\\
\hspace{-0.2cm}\mathbf{0}_{\bar{o}},\hspace{0.86cm}t \in \left[ {0,\rho } \right)
\end{array} \right.\label{eq:asg}
\end{align}
in system~(\ref{eq:sl1}) is a \emph{zero-dynamics attack}, if $\mathbf{0}_{\bar{n}} \neq \tilde{z}(0) - z(0) \in \mathbb{R}^{\bar{n}}$, $\mathbf{0}_{\bar{o}} \neq g(\rho) \in \mathbb{R}^{\bar{o}}$, $\rho \geq 0$ and $\eta \in \mathbb{C}$ satisfy
\begin{subequations}
\begin{align}
\!\!\tilde{z}\left( 0 \right) - z\left( 0 \right) &\in \ker \left( {{\mathcal{O}}} \right), \hspace{0.1cm} \emph{\emph{if}} \hspace{0.1cm}  \rho > 0\label{eq:aczeo}\\
\!\!\left[\!
    \begin{array}{c}
        {{e^{A\rho }}\left( {\tilde{z}\left( 0 \right) - z\left( 0 \right)} \right)}\\ \hdashline[2pt/2pt]
        { - g\left( \rho  \right)}
    \end{array}
\!\right] &\in \ker \left(\left[\!
    \begin{array}{c;{2pt/2pt}c}
        {\eta \mathbf{1}_{\bar{n} \times \bar{n}} - A} & B\\ \hdashline[2pt/2pt]
        -C & D
    \end{array}
\!\right]\right),\label{eq:czeo}
\end{align}\label{eq:ssczeo}
\end{subequations}where \begin{align}
{\mathcal{O}} \triangleq \left[
    \begin{array}{c;{1pt/1pt}c;{1pt/1pt}c;{1pt/1pt}c}
        {C^\top} & {{\left( {CA} \right)}^\top} & \ldots & {{\left( {C{A^{\bar{n} - 1}}} \right)}^\top}
    \end{array}
\right]^\top.\label{eq:1aczeo}
\end{align}
\label{thm:zods}
\end{definition}
\begin{cor}
Under the ZDA~(\ref{eq:ssczeo}), the states and monitored outputs of systems~(\ref{eq:sl1}) and~(\ref{eq:sl2}) satisfy
\begin{align}
y\left( t \right) &= \tilde y\left( t \right),\hspace{0.1cm} \emph{\emph{for}} \hspace{0.1cm} \emph{\emph{all}} \hspace{0.1cm} t \ge 0 \label{eq:rs1}\\
\tilde{z}\left( t \right) &= \left\{ \begin{array}{l}
\hspace{-0.2cm}{e^{At}}\tilde{z}\left( 0 \right), \hspace{2.98cm}t \in \left[ {0,\rho } \right]\\
\hspace{-0.2cm}z\left( t \right) + \left( {\tilde{z}\left( \rho  \right) - z\left( \rho  \right)} \right){e^{\eta \left( {t - \rho } \right)}}, t \in \left( {\rho,\infty } \right).
\end{array} \right.\label{eq:rs2}
\end{align}\label{thm:rezda}
\end{cor}
\begin{IEEEproof}
See Appendix A.
\end{IEEEproof}

\begin{rem}
The state solution~(\ref{eq:rs2}) shows that through choosing the parameter $\eta$ and also the attack-starting time $\rho$, the attacker can achieve various objectives, see e.g.,
 \begin{itemize}
\item $\rho = \infty$: altering the steady-state value while not affecting system stability;
 \item $\rho < \infty$, ${\mathop{\rm Re}\nolimits} \left( \eta  \right)$ $>$ $0$: making system unstable;
 \item $\rho$ $<$ $\infty$, ${\mathop{\rm Re}\nolimits} \left( \eta  \right)$ $=$ $0$, ${\mathop{\rm Im}\nolimits} \left( \eta  \right)$ $\neq$ $0$: causing oscillatory behavior.
 \end{itemize}
The output~(\ref{eq:rs1}) indicates the undetectable/stealthy property of proposed ZDA~(\ref{eq:asg}).
\end{rem}

\begin{rem}[Mixed Stealthy Attacks]
To launch the ZDA, the attacker must modify initial condition; otherwise, $\tilde{z}\left( 0 \right)-z\left( 0 \right)  = {\mathbf{0}_{\bar{n}}}$, $e^{A\rho}(\tilde{z}\left( 0 \right)-z\left( 0 \right))  = {\mathbf{0}_{\bar{n}}}$, which with~(\ref{eq:czeo}) implies that $Bg(\rho) = {\mathbf{0}_{\bar{n}}}$ and $Dg(\rho) = {\mathbf{0}_{\bar{m}}}$. Thus, the attack signal~(\ref{eq:asg}) does not have any effect on the system~(\ref{eq:sl1}). The attack policy~(\ref{eq:ssczeo}) in conjunction with the property~(\ref{eq:rs1}) implies that in the situation where $\rho > 0$, i.e., the attack-starting time is not the initial time,  the attack strategy comprises two stealthy attacks, which can be well illustrated in the example of cyber-physcial systems:
\begin{itemize}
  \item Before the starting time $\rho$, the attacker injects false data to the data of initial condition sent to the Luenberger observer (attack detector~\cite{F13}) in cyber layer, while keeping stealthy, i.e., $y\left( t \right) = \tilde y\left( t \right)$ for $t \in[0, \rho)$.
  \item At the starting time $\rho$, the attacker introduces signals of ZDA $g(t) = g{e^{\eta \left( {t - \rho } \right)}}, t \geq \rho,$ to the system.
\end{itemize}\label{thm:hho}
\end{rem}

\section{Problem Formulation}
For simplicity, we let the increasingly ordered set $\mathbb{M} \triangleq \{1, 2, \ldots\}$ $\subseteq \mathbb{V}$ denote the set of monitored agents.

\subsection{System in The Absence of Attacks}
Under the simplified control protocol proposed in~\cite{YA}, the second-order multi-agent system with monitored outputs is described by
\begin{subequations}
\begin{align}
{\dot x_i}\left( t \right) &= {v_i}\left( t \right)\label{eq:oonp1}\\
{{\dot v}_i}\left( t \right) &= \sum\limits_{j = 1}^n {a_{ij}^{{\sigma} \left( t \right)}\!\left(
{{x_j}\left( t \right) - {x_i}\left( t \right)} \right)}, \hspace{0.2cm}i \in \mathbb{V}\label{eq:oonp2}\\
{{y}_j}\left( t \right) &= {x}_{j}(t)\!, j \in \mathbb{M}\label{eq:oonp3}
\end{align}\label{eq:oonp}\end{subequations}
where $x_{i}(t) \in \mathbb{R}$ is the position, $v_{i}(t) \in \mathbb{R}$ is the velocity, ${y_j}\left( t\right) \in \mathbb{R}$ is the output of monitored agent $i$ used to detect stealthy attack, ${\sigma}(t):$ $[0,\infty ) \to \mathfrak{S} \triangleq \{1,2,\ldots,s\}$, is the topology-switching signal. Here, ${\sigma} (t) = {p_k} \in \mathfrak{S}$ for $t \in [t_{k}, t_{k+1})$ means the $p^{\emph{\emph{th}}}$ topology is activated over time interval $[t_{k}, t_{k+1}), k \in \mathbb{N}_{0}$, and $a^{p_{k}}_{ij}$ is the entry of the weighted adjacency matrix which describes the activated $p^{\emph{\emph{th}}}$ topology of communication network.

\subsection{System in The Presence of Attacks}
We usually refer to an agent under attack as a \emph{misbehaving agent}~\cite{n12}. We let $\mathbb{K}$ $\subseteq$ $\mathbb{V}$ denote the set of misbehaving agents. The multi-agent system~(\ref{eq:oonp}) under ZDA is described by
\begin{subequations}
\begin{align}
\!\!&{{\dot{\tilde{x}}}_i}\left( t \right) = {{\tilde v}_i}\left( t \right) \label{eq:oox1}\\
\!\!&{{\dot{\tilde v}}_i}\left( t \right) = \sum\limits_{i = 1}^n\! {a_{ij}^{\tilde {\sigma}\left( t \right)}\!\left( {{{\tilde x}_j}\left( t \right) - {{\tilde x}_i}\left( t \right)} \right)}  + \left\{ \begin{array}{l}
\hspace{-0.2cm}{\tilde{g}}_i\!\left( t \right)\!, i \in \mathbb{K}\\
\hspace{-0.2cm}0, \hspace{0.10cm} i \in {\mathbb{V}} \backslash \mathbb{K}
\end{array} \right.\label{eq:oox2}\\
\!\!&{{\tilde{y}}_j}\left( t \right) = \tilde{x}_{j}(t), j \in \mathbb{M}\label{eq:oox3}
\end{align}\label{eq:oofn}\end{subequations}
where ${g_i}\left( t\right)$ is the ZDA signal in the form of~(\ref{eq:asg}):
\begin{align}
{\tilde{g}_i}\left( t \right) = \left\{ \begin{array}{l}
\hspace{-0.2cm}{g_i}{e^{\eta \left( {t - \rho } \right)}},  t \in \left[ {\rho ,\infty } \right)\\
\hspace{-0.2cm}0,\hspace{1.15cm}\emph{\emph{otherwise}}.
\end{array} \right.\label{eq:nm2a}
\end{align}

We note that system~(\ref{eq:oofn}) can equivalently transforms to a switched system under attack:
\begin{subequations}
\begin{align}
\dot{\tilde z}\left( t \right) &= {A_{\tilde \sigma \left( t \right)}}\tilde z\left( t \right) + g\left( t \right)\\
\tilde y\left( t \right) &= C\tilde z\left( t \right)
\end{align}\label{eq:s1}\end{subequations}
where we define the following variables and matrices:
\begin{subequations}
\begin{align}
\!\!\!A_{\tilde{\sigma}(t)} &\triangleq \left[
    \begin{array}{c;{2pt/2pt}c}
        \mathbf{0}_{n \times n} & \mathbf{1}_{n \times n}\\ \hdashline[2pt/2pt]
        -\mathcal{L}_{\tilde{\sigma}(t)} & \mathbf{0}_{n \times n}
    \end{array}
\right],\label{eq:nm0}\\
C &\triangleq\left[\!\!
    \begin{array}{c;{1pt/1pt}c;{1pt/1pt}c;{1pt/1pt}c}
        {\mathrm{e}_1} & \ldots & {\mathrm{e}_{\left| \mathbb{M} \right|}} & {\mathbf{0}_{\left| \mathbb{M} \right| \times 2n - \left| \mathbb{M} \right|}}
    \end{array}
\!\!\!\right]^\top, \label{eq:nm0a1}\\
\!\!\!\!\tilde{z}\left( t \right) &\triangleq\left[\!\!
    \begin{array}{c;{1pt/1pt}c;{1pt/1pt}c;{1pt/1pt}c;{1pt/1pt}c;{1pt/1pt}c}
        {\tilde{x}_1}\left( t \right) \!&\! \ldots \!&\! {\tilde{x}_n}\left( t \right) \!&\! {\tilde{v}_1}\left( t \right) \!&\! \ldots \!&\! {\tilde{v}_n}\left( t \right)
    \end{array}
\!\!\!\right]^\top,\label{eq:ssd1}\\
\!\!\!g\left( t \right) &\triangleq \left[\!\!
    \begin{array}{c;{1pt/1pt}c}
        \mathbf{0}_n^\top & a^{\top}\left( t \right)
    \end{array}
\!\!\right]^\top\!,\label{eq:ndm4pp}\\
\!\!\![{a}\left( t \right)]_{i} &\!\triangleq\! \left\{ \begin{array}{l}
\hspace{-0.2cm}{\tilde{g}_i}\left( t \right),i \in \mathbb{K}\\
\hspace{-0.2cm}0, \hspace{0.62cm}i \in {\mathbb{V}} \backslash \mathbb{K}
\end{array} \right.\label{eq:ndm4}
\end{align}
\end{subequations}
where $\mathrm{e}_i \in \mathbb{R}^{n}$ is the $i^{\emph{\emph{th}}}$ vector of the canonical basis.
In addition, system~(\ref{eq:oonp}) equivalently transforms to a switched system:
\begin{subequations}
\begin{align}
\dot{z}\left( t \right) &= {A_{\sigma \left( t \right)}}z\left( t \right)\\
y\left( t \right) &= Cz\left( t \right).
\end{align}\label{eq:s2}\end{subequations}

We present the definition of time-dependent switching, which is part of our defense strategy.
\begin{definition}
The topology-switching signals $\sigma(t)$ and  $\tilde{\sigma}(t)$ in multi-agent systems~(\ref{eq:s1}) and~(\ref{eq:s2}) are said to be time-dependent if the switching times and topologies depend only on time, such that
\begin{align}
\sigma(t) = \tilde{\sigma}(t), \hspace{0.1cm} \emph{\emph{for}} \hspace{0.1cm} \emph{\emph{all}} \hspace{0.1cm} t \geq 0. \label{eq:tde}
\end{align}
\label{thm:td}
\end{definition}

We made the following assumptions on the attacker and defender.
\begin{asm}
The attacker
\begin{enumerate}
  \item knows the currently activated topology and its dwell time;
  \item has the memory of the past switching sequences.
\end{enumerate}\label{thm:att}
\end{asm}
\begin{asm}
The defender
\begin{enumerate}
  \item designs the switching sequences including switching times and topologies;
  \item  has no knowledge of the attack-starting time;
  \item  has no knowledge of the number of misbehaving agents.
\end{enumerate} \label{thm:auatt}
\end{asm}

\subsection{Attack-Starting Time}
The attack-starting time $\rho$ of the signal~(\ref{eq:asg}) plays a critical role in guaranteeing the stealthy property~(\ref{eq:rs1}), which can be utilized by the attacker to escape from being detected by the defense strategy of modifying input or output matrix~\cite{T12} finitely over finite time. If the attack-starting time is not reasonable, the changes in system dynamics induced by attacker's action ``start attack" at $\rho$ can be used by defender to detect the stealthy attack. Therefore, from the perspective of stealthy attack design, it is not trivial to study how the attacker should use its knowledge and memory to decide the attack-starting time to guarantee its stealthy.

Before presenting the selection scheme of attack-starting time, let us define:
\begin{subequations}
 \begin{align}
 \breve{z}\!\left( t \right) &\triangleq \tilde{z}\!\left( t \right) \!-\! z\!\left( t \right) \!=\! \left[\!\begin{array}{c}
        {\tilde{x}}\!\left( t \right)\\ \hdashline[2pt/2pt]
         {\tilde{v}}\!\left( t \right)
    \end{array}
\!\right] \!-\! \left[\!\!\begin{array}{c}
        {{x}}\!\left( t \right)\\ \hdashline[2pt/2pt]
         {{v}}\!\left( t \right)
    \end{array}
\!\right] \!=\! \left[\!\!\begin{array}{c}
        {\breve{x}}\!\left( t \right)\\ \hdashline[2pt/2pt]
         {\breve{v}}\!\left( t \right)
    \end{array}
\!\right], \label{eq:krs1}\\
{\breve{y}}\left( t \right) &\triangleq \tilde y\left( t \right) - y\left( t \right),\label{eq:krs2}\\
{P_{{\sigma} \left( {{t_k}} \right)}} &\triangleq \left[
    \begin{array}{c;{2pt/2pt}c}
        {\eta {{\bf{1}}_{2n \times 2n}} - {A_{{\sigma} \left( {{t_k}} \right)}}} & \mathbf{1}_{2n \times 2n}\\ \hdashline[2pt/2pt]
        -C & {{{\mathbf{0}_{\left| \mathbb{M} \right| \times 2n}}}}
    \end{array}
\right], \label{eq:pal21}\\
\breve{z}\left( t \right) &\triangleq {e^{{A_{\sigma \left( {{t_k}} \right)}}\left( {t - {t_k}} \right)}}\prod\limits_{o = 0}^{k - 1} {{e^{{A_{\sigma \left( {{t_o}} \right)}}\left( {{t_{o + 1}} - {t_o}} \right)}}}\breve{z}\left( 0 \right), \label{eq:pal22}\\
g &\triangleq \left[
    \begin{array}{c;{1pt/1pt}c;{1pt/1pt}c;{1pt/1pt}c}
        \mathbf{0}_n^\top & {g_1} & \ldots & {g_{n}}
    \end{array}\right],\label{eq:pal23}\\
\mathcal{O}_{k} &\triangleq \left[\!\!\!
    \begin{array}{c;{1pt/1pt}c;{1pt/1pt}c;{1pt/1pt}c}
        {C^\top} & {{{\left( {{C}{A_{\sigma \left( {{t_k}} \right)}}} \right)}^\top}} & \ldots & {{{\left( {{C}A_{\sigma \left( {{t_k}} \right)}^{2n - 1}} \right)}^\top}} \end{array} \!\!\!\right]^\top\!\!. \label{eq:pal24}\\
        \mathbf{N}^{k}_{k} &= \ker\left ( {\mathcal{O}}_{m} \right ), \label{eq:hhp1}\\
\mathbf{N}^{k}_{q} &= \ker({\mathcal{O}}_{q} ) \bigcap e^{-A_{\sigma(t_{q})}\tau_{q}}\mathbf{N}^{k}_{q+1}, 1 \leq q \leq k-1. \label{eq:hhp1aa}
\end{align}\label{eq:ppaa}
\end{subequations}

\begin{algorithm}
  \caption{Attack-Starting Time $\rho$}
  \KwIn{Sets: $\mathbf{N}^{k}_{1}$ recursively computed by~(\ref{eq:hhp1}) and~(\ref{eq:hhp1aa}), and 
  \begin{align}
\!\!{\mathcal{S}_{k}} \!\triangleq\! \!\left\{ \!{t\!: \!\!
\left[\!\!\begin{array}{c}
        \breve{z}\left( t \right)\\ \hdashline[2pt/2pt]
         -g
    \end{array}
\!\right] \!\!\in\! \ker \!\left(\! {{P_{{\sigma} \left(\! {{t_k}} \!\right)}}} \!\right)}, t \!\in\! [t_{k}, \!t_{k\!+\!1}]\!\right\}\label{eq:hhp2}\end{align}
with ${P_{\sigma(t_{k})}}$, $\breve{z}\left( t \right)$, $g$ and ${\mathcal{O}_{k}}$ defined in~(\ref{eq:ppaa}).
  }
    \eIf{$\breve{z}\left( 0 \right) \not\in \mathbf{N}^{k}_{1}$}
    {Choose $\rho = t_{k}$ at current $t_{k}$;}
    {\eIf{$\mathop {\max }\limits_{t \in {\mathcal{S}_{k}}} \left\{ t \right\} \ne {t_{k + 1}}$ }
    {Choose $\rho \in \mathcal{S}_{k}$ at current $t_{k}$;
    }
    {Choose $\rho$ at current $t_{k}$ or next $t_{k+1}$. } }
\end{algorithm}

\begin{prop}
Under time-dependent topology switching~(\ref{eq:tde}), the action ``start attack" of ZDA in the system~(\ref{eq:s1})
does not affect the stealthy property~(\ref{eq:rs1})  if and only if the attack-starting time $\rho$ is generated by Algorithm~1.
\label{thm:asmsaa}
\end{prop}

\begin{IEEEproof}
See Appendix B.
\end{IEEEproof}

\subsection{Strategy on Switching Times}
Inspired by~\cite{T12}, the core of defense strategy proposed in this paper is to make changes on system dynamics through changing communication topology such that the attack policy~(\ref{eq:ssczeo}) is not feasible. In the realistic situation where the defender has no knowledge of the attack-starting times, to detect ZDA we must consider infinitely changing topology over infinite time. The strategy on switching times described by the following lemma, which are studied in Part I paper and will also be used for observer design in this Part II paper, addresses the problem: \emph{when should the topology of multi-agent system~(\ref{eq:oonp}) switch to detect ZDA, such that the changes in the system dynamics do not destroy system stability in the absence of attacks?}
\begin{lem}~\cite{YA}
Consider the second-order multi-agent system~(\ref{eq:oonp}). For the given topology set $\mathfrak{S}$ that satisfies
\begin{align}
&\forall r \in \mathfrak{S}: \sqrt {\frac{{{\lambda _i}\left( {{{\cal L}_r}} \right)}}{{{\lambda _j}\left( {{{\cal L}_r}} \right)}}} \in \mathbb{Q}, \hspace{0.1cm} \emph{\emph{for}} \hspace{0.1cm} \forall i,j = 2, \ldots ,n\label{eq:bss1}
\end{align}
and the scalars $1 > \beta > 0$, $\alpha > 0$ and $\kappa \in \mathbb{N}$, if the dwell times $\tau_{r}$, $r \in \mathfrak{S}$, satisfy
\begin{align}
\tau_{r} = {{\widehat \tau_{\max} }} + \mathrm{m}\frac{{T}_{r}}{2}, \mathrm{m} \in \mathbb{N} \label{eq:addt}
\end{align}
where $0 < {{\widehat \tau_{\max} }} < \frac{{ - \ln \beta }}{\alpha }$, $0 < {{\widehat \tau_{\max} }} + {\frac{\mathrm{m}{T}_{r}}{2}} - \left( {{\beta ^{ - \frac{1}{\kappa}}} - 1} \right)\frac{\kappa}{{\alpha  - \xi }}$, $\xi < \alpha$, $\xi = \mathop {\max }\limits_{r \in \mathfrak{S},i = 1, \ldots ,n} \left\{ {1 - {\lambda_i}\left( {{{\cal L}_r}} \right)}, {-1 + {\lambda_i}\left( {{{\cal L}_r}} \right)} \right\}$ and ${T}_{r} = \emph{\emph{lcm}}\left ( \frac{2\pi }{\sqrt{{\lambda_i}(\mathcal{L}_r})}; i = 2, ..., n \right )$, then the asymptotic second-order consensus is achieved.
\label{thm:lmr}
\end{lem}

\begin{rem}
Let us assume that at time $t^{-}_{k}$, system~(\ref{eq:s1}) or~(\ref{eq:s2}) is already at the steady state. It verifies that under the attack signal~(\ref{eq:asg}), at the topology-switching time $t_{k}$, $x(t_{k})$ = $x(t^{-}_{k})$ and $v(t_{k})$ = $v(t^{-}_{k})$, and the system maintains its steady state at $t_{k}$, which means topology switching does not have impulsive effect on the systems~(\ref{eq:s1}) and~(\ref{eq:s2}). We should note that the defense strategy (strategic topology switching) that will be developed in the following sections cannot be directly applied to such multi-agent systems that topology switching has impulsive effect. \end{rem}

\section{Detectability of Zero-Dynamics Attack}
This section focuses on the detectability of ZDA, which will answer the question: \emph{what topologies of multi-agent system~(\ref{eq:oonp}) to switch to such that the attack policy~(\ref{eq:ssczeo}) is not feasible?} To better illustrate the strategy on switching topologies, we introduce the definitions of components in a graph and difference graph.
\begin{definition}[Components of Graph~\cite{NW10}] The \emph{components} of a graph are its maximal connected subgraphs. A component is said to be \emph{trivial} if it has no edges; otherwise, it is a \emph{nontrivial component}.
\end{definition}
\begin{definition}
The difference graph $\mathrm{G}^{rs}_{\emph{\emph{diff}}} = \left( {\mathbb{V}_{\emph{\emph{diff}}}^{rs},\mathbb{E}_{\emph{\emph{diff}}}^{rs}} \right)$ of two graphs $\mathrm{G}_{r}$ and $\mathrm{G}_{s}$ is generated as 
\begin{align}
\mathbb{V}_{\emph{\emph{diff}}}^{rs} &= {\mathbb{V}_r} \cup {\mathbb{V}_s} \nonumber\\
\left( {i,j} \right) &\in \mathbb{E}_{\emph{\emph{diff}}}^{rs}, \hspace{0.1cm}\emph{\emph{if}} \hspace{0.1cm} a_{ij}^r - a_{ij}^s \ne 0 \nonumber
\end{align}
where $\mathbb{V}_r$ and $a^{r}_{ij}$ are the set of vertices (agents) and the entry of weighted adjacency matrix of the graph $\mathrm{G}_{r}$, respectively. 
\end{definition}

We define the union difference graph for switching difference graphs as:
\begin{align}
{\mathrm{G}_{\emph{\emph{diff}}}} &\triangleq \left( {\bigcup\limits_{r,s \in \mathfrak{S}} {\mathbb{V}_{\emph{\emph{diff}}}^{rs}} ,\bigcup\limits_{r,s \in \mathfrak{S}} {\mathbb{E}_{\emph{\emph{diff}}}^{rs}} } \right).\label{eq:udg}
\end{align}

\begin{figure}
\centering
\includegraphics[scale=0.6]{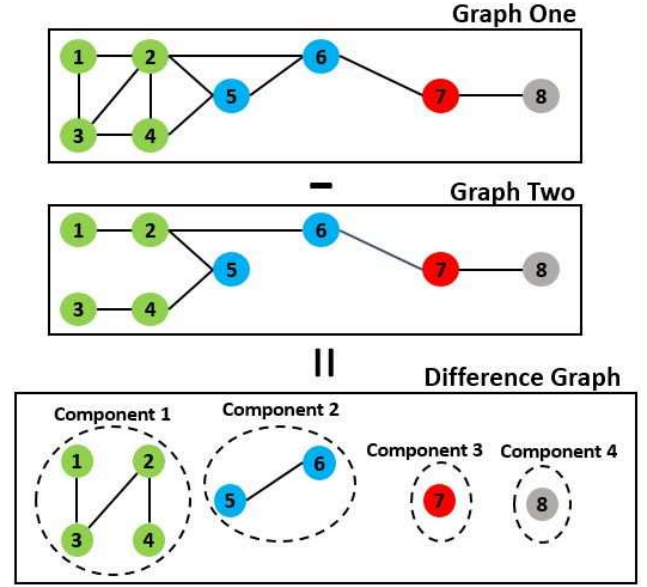}
\caption{Components of difference graph (the weights of communications links are uniformly set as ones).}
\label{fig:cpp}
\end{figure}

We use $\mathbb{C}_{i}({\mathrm{G}_{\emph{\emph{diff}}}})$ to denote the set of agents in $i^{\emph{\emph{th}}}$ component of union difference graph ${\mathrm{G}_{\emph{\emph{diff}}}}$. Obviously, $\mathbb{V}$ $=$ $\mathbb{C}_{1}$ $\bigcup$ $\mathbb{C}_{2}$ $\bigcup$ $\ldots$ $\bigcup$ $\mathbb{C}_{\mathrm{d}}$, and $\mathbb{C}_{\mathrm{p}}$ $\bigcap$ $\mathbb{C}_{\mathrm{q}}$ $=$ $\emptyset$ if $\mathrm{p}$ $\neq$ $\mathrm{q}$, where $\mathrm d$ is the number of the component of graph ${\mathrm{G}_{\emph{\emph{diff}}}}$. As an example, the difference graph in Figure~\ref{fig:cpp} has two nontrivial components, $\mathbb{C}_{1}({\mathrm{G}_{\emph{\emph{diff}}}}) = \left\{ {{1},{2}, {3},{4}} \right\}$, $\mathbb{C}_{2}({\mathrm{G}_{\emph{\emph{diff}}}}) = \left\{ {{5},{6}} \right\}$, and two trivial components, $\mathbb{C}_{3}({\mathrm{G}_{\emph{\emph{diff}}}}) = \left\{ {{7}} \right\}$, $\mathbb{C}_{4}({\mathrm{G}_{\emph{\emph{diff}}}}) = \left\{ {{8}} \right\}$.

In the following theorem, we present the strategy on switching topologies.

\begin{thm} Consider the multi-agent system~(\ref{eq:s1}). Under time-dependent topology switching, the ZDA can be detected by defender without knowledge of the numbder of misbehaving agents and the attack-starting time, if and only if each component of union difference graph has at least one monitored agent, i.e.,
  \begin{align}
  {\mathbb{C}_i}({\mathrm{G}_{\emph{\emph{diff}}}}) \cap \mathbb{M} \ne \emptyset ,\forall i = 1, \ldots \mathrm{d}.\label{eq:cccc}
  \end{align}
\label{thm:sms}
\end{thm}

\begin{IEEEproof}
Under result of time-dependent topology switching~(\ref{eq:tde}), from~(\ref{eq:s1}) and~(\ref{eq:s2}), we have:
\begin{subequations}
\begin{align}
\!\!\!&\dot{\breve{z}}\left( t \right) = {A_{\sigma (t)}}\breve{z}\left( t \right) + {g}\left( t\right),\\
\!\!\!&{{\breve{y}}}\left( t \right) = C\breve{z}\left( t \right),
\end{align}\label{eq:us}\end{subequations}
where $\breve{z}\left( t \right)$ and ${\breve{y}}\left( t \right)$ are given by~(\ref{eq:krs1}) and~(\ref{eq:krs2}), respectively.

Let us define:
\begin{align}
{\mathcal{P}_r} = \left[
    \begin{array}{c;{1pt/1pt}c}
       \!\!\! {\eta{\mathbf{1}_{2n \times 2n}} - {A_{r}}}\!\!\! & \!\!\!\mathbf{1}_{2n \times 2n}\!\!\! \\ \hdashline[1pt/1pt]
       \!\!\! {-C}\!\!\! & \!\!{\mathbf{0}_{{\left| \mathbb{M} \right|} \times 2n}\!\!}
    \end{array}
\right]. \label{eq:cmo}
\end{align}

\textbf{(Sufficient Condition)} We now assume to the contrary that the system~(\ref{eq:s1}) is under ZDA. By Definition~\ref{thm:zods}, we obtain
\begin{align}
\left[\!
    \begin{array}{c}
        \!\!\!\breve{z}(\rho)\!\!\! \\ \hdashline[1pt/1pt]
        \!\!\!-{g}(\rho)\!\!\!
    \end{array}
\!\right] \in \ker \!\left(  {\mathcal{P}_{\sigma(\rho)}} \right).\label{eq:us1pp}
\end{align}

By the resulted state~(\ref{eq:rs2}) and the form of ZDA signal~(\ref{eq:asg}), we obtain that
\begin{align}
\left[ {\breve{z}\left( t \right), -{g}(t)} \right] = {e^{\eta (t - \rho )}}\left[ {\breve{z}\left( \rho  \right), -{g}(\rho )} \right],t \ge \rho. \nonumber
\end{align}
Since ${e^{{\eta (t} - \rho) }} \neq 0$, we conclude that system~(\ref{eq:s1}) has ZDA is equivalent to
\begin{align}
\left[\!
    \begin{array}{c}
        \!\!\!\breve{z}(\rho)\!\!\! \\ \hdashline[1pt/1pt]
        \!\!\!-{g}(\rho)\!\!\!
    \end{array}
\!\right] \in \bigcap\limits_{r \in \mathfrak{S}} {\ker \left( {{\mathcal{P}_r}} \right)}.\label{eq:us1}
\end{align}

Substituting $C$ in~(\ref{eq:nm0a1}),  $\breve{z}\left( \rho \right)$ in~(\ref{eq:krs1}), $g(\rho)$ in~(\ref{eq:ndm4pp}) with~(\ref{eq:ndm4}), $A_{r}$ in~(\ref{eq:nm0}), and $\mathcal{P}_{r}$ in~(\ref{eq:cmo}) into~(\ref{eq:us1}) and expanding it out yields
\begin{align}
\!\!\!\left[\!
    \begin{array}{c;{1.0pt/1.0pt}c;{1.0pt/1.0pt}c;{1.0pt/1.0pt}c}
        \!\!{\eta{\mathbf{1}_{n\times n}}}\!\!\!\! & \!\!\!-{\mathbf{1}_{n\times n}}\!\!\!\! & \!\!{\mathbf{0}_{n \times n}}\!\!\!\! & \!\!{\mathbf{0}_{n \times n}}\!\!\!\!  \\ \hdashline[1pt/1pt]
        \!\!\!\!{ \mathcal{L}_{r}}\!\!\!\! & \!\!\!{\eta{\mathbf{1}_{n\times n}}}\!\!\!\! & \!\!{\mathbf{0}_{n \times n}}\!\!\!\! & \!\!{\mathbf{1}_{n \times n}}\!\!\!\! \\ \hdashline[1pt/1pt]
        \!\!\!\!-{\mathbf{e}}^{\top}_{j}\!\!\!\! & \!\!\!\mathbf{0}^{\top}_{n}\!\!\!\! & \!\!{\mathbf{0}^{\top}_{n}}\!\!\!\! & \!\!\mathbf{0}^{\top}_{n}\!\!\!\!
    \end{array}
\!\right]\!\!
\left[
    \begin{array}{c}
        \!\!\!\breve{x}(\rho)\!\!\!\! \\ \hdashline[1pt/1pt]
        \!\!\!\breve{ v}(\rho)\!\!\!\! \\ \hdashline[1pt/1pt]
        \!\!\!\mathbf{0}_{n}\!\!\!\! \\ \hdashline[1pt/1pt]
        \!\!\!{-{a}}\left( \rho \right)\!\!\!\!
    \end{array}
\right] \!\!=\!\!\left[
    \begin{array}{c}
        \!\!\mathbf{0}_{n}\!\!\! \\ \hdashline[1pt/1pt]
        \!\!\mathbf{0}_{n}\!\!\! \\ \hdashline[1pt/1pt]
        \!\!0\!\!\!
    \end{array}
\right], \!\forall j \!\in\! \mathbb{M}, \!\forall r \!\in\! \mathfrak{S} \nonumber
\end{align}
which is equivalent to
\begin{align}
\eta \breve{x}\left( \rho \right) - \breve{v}\left( \rho \right) &= \mathbf{0}_{n},\label{eq:ss1}\\
-{a}\left( \rho \right) + {\mathcal{L}_r}\breve{x}\left( \rho \right) + \eta \breve{v}(\rho) &= \mathbf{0}_{n}, \!\forall r \in \mathfrak{S}\label{eq:ss2}\\
-{a}\left( \rho \right) + {\mathcal{L}_s}\breve{x}\left( \rho\right) + \eta \breve{v}(\rho) &= \mathbf{0}_{n}, \!\forall s \in \mathfrak{S}\label{eq:ss3}\\
{\breve{x}_j}\left( \rho \right) &= 0, \forall j \in \mathbb{M}.\label{eq:ss4}
\end{align}

Through elementary row transformation, the Laplacian matrix of union difference graph ${\mathrm{G}_{\emph{\emph{diff}}}}$ can be written as
\begin{equation}
\tilde{\mathcal{L}} \triangleq \emph{\emph{diag}}\left\{ {\mathcal{L}\left( {{\mathbb{C}_1}}({\mathrm{G}_{\emph{\emph{diff}}}}) \right), \ldots ,\mathcal{L}\left( {{\mathbb{C}_\mathrm{d}}}({\mathrm{G}_{\emph{\emph{diff}}}}) \right)} \right\},\label{eq:aas}
\end{equation}
where $\mathcal{L}\left( {{\mathbb{C}_\mathrm{q}}}({\mathrm{G}_{\emph{\emph{diff}}}}) \right)$, $q \in \{1,2,\ldots,\mathrm{d}\}$, denote the Laplacian matrix of the $\mathrm{q}^{\emph{\emph{th}}}$ component.

Noting that equation (\ref{eq:ss2}) subtracting equation~(\ref{eq:ss3}) results in
$\left( {{\mathcal{L}_r} - {\mathcal{L}_s}} \right)\breve{x}\left( \kappa \right) = \mathbf{0}_{n}, \forall r, s \in \mathfrak{S}$, which is equivalent to 
\begin{align}
\tilde{\mathcal{L}}\breve{x}\left( \kappa \right) = \mathbf{0}_{n},\label{eq:fsp}
\end{align}
where $\tilde{\mathcal{L}}$ is defined in (\ref{eq:aas}). From~\cite{ada11}, it is known that the Laplacian matrix 
of component $\mathcal{L}\left( {{\mathbb{C}_\mathrm{q}}}({\mathrm{G}_{\emph{\emph{diff}}}}) \right)$ has properties: i) zero is one of its eigenvalues with multiplicity one, ii) the eigenvector that corresponds to the eigenvalue zero is $\textbf{1}_{| \mathbb{C}_{q}({\mathrm{G}_{\emph{\emph{diff}}}}) |}$, $\forall \mathrm{q} \in \{1, \ldots, \mathrm{d}\}$. It follows from~(\ref{eq:ss4}) and~(\ref{eq:cccc}) that the solution of~(\ref{eq:fsp}) is obtained as ${\breve{x}}\left( \rho \right) =  \mathbf{0}_{n}$, which works with~(\ref{eq:ss1}) implies that ${\breve{v}}\left( \rho \right) =  \mathbf{0}_{n}$, substituting which into~(\ref{eq:ss2}) or~(\ref{eq:ss3})  yields the same result as $a\left( \rho\right) = \mathbf{0}_{n}$, which, in conjunction with~(\ref{eq:ndm4pp}), implies $g\left( \rho \right) = \mathbf{0}_{2n}$, indicating there is no ZDA by Definition~\ref{thm:zods}. Thus, a contradiction occurs.

\textbf{(Necessary Condition)} Substituting (\ref{eq:ss1}) into (\ref{eq:ss2}) yields $\left( {{{\cal L}_r} + {\eta ^2}{\mathbf{1}_{n \times n}}} \right)\breve{x}\left( \rho  \right) = a\left( \rho  \right), \forall r \in \mathfrak{S}$, which is equivalent to
\begin{subequations}
\begin{align}
&({\sum\limits_{r = 1}^{\left| \mathfrak{S} \right|} {{\alpha _r}{\mathcal{L}_r}}  + {\eta ^2}{{\bf{1}}_{n \times n}}})\breve{x}\left( \rho  \right) = a\left( \rho  \right),\label{eq:mmp}\\
&\forall {\alpha _r} > 0,\sum\limits_{r = 1}^{\left| \mathfrak{S} \right|} {{\alpha _r} = 1}.
\end{align}\label{eq:ss4cd}\end{subequations}
If ${\mathop{\rm Im}\nolimits} \left( \eta  \right) \ne 0$, (\ref{eq:ss1}) shows that \begin{align}
\exists i \in \mathbb{V}: {\mathop{\rm Im}\nolimits} \left( {{\breve{x}_i}\left( \rho  \right)} \right) \ne 0 \hspace{0.1cm} \emph{\emph{or}} \hspace{0.1cm} {\mathop{\rm Im}\nolimits} \left( {{\breve{v}_i}\left( \rho  \right)} \right) \ne 0, \nonumber
\end{align}
which contradicts with  $\breve{v}\left( \rho \right) \in \mathbb{R}^{n}$ and $\breve{x}\left( \rho \right) \in \mathbb{R}^{n}$ in the definition of ZDA. Therefore, in the following proof, we need to consider only the cases of $({\mathop{\rm Im}\nolimits} \left( \eta  \right) = 0$, ${\mathop{\rm Re}\nolimits} \left( \eta  \right) = 0)$ and $({\mathop{\rm Im}\nolimits} \left( \eta  \right) = 0$, ${\mathop{\rm Re}\nolimits} \left( \eta  \right) \ne 0)$.

Case One--$({\mathop{\rm Im}\nolimits} \left( \eta  \right) = 0$,  ${\mathop{\rm Re}\nolimits} \left( \eta  \right) \ne 0)$: We note that there exists one implied condition that is the union graph $\mathrm{G} \triangleq \left( {\bigcup\limits_{r \in \mathfrak{S}} {{\mathbb{V}_r}} ,\bigcup\limits_{r \in \mathfrak{S}} {{\mathbb{E}_r}} } \right)$ of switching topologies in $\mathfrak{S}$ is connected; if not, the asymptotic second-order consensus cannot be achieved, which is undesirable. It is straightforward to verify that if the condition (\ref{eq:cccc}) is not satisfied, the equation  (\ref{eq:fsp}) has the solution that has non-identical entries.  Moreover, if $\eta \neq 0$, ${\sum\limits_{r = 1}^{\left| \mathfrak{S} \right|} {{\alpha _r}{\mathcal{L}_r}}  + {\eta ^2}{{\bf{1}}_{n \times n}}}$ is full-rank for $\forall {\alpha _r} > 0,\sum\limits_{r = 1}^{\left| \mathfrak{S} \right|} {{\alpha _r} = 1}$. Thus, we obtain a feasible non-zero vector ${a}\left( \rho \right)$ 
from (\ref{eq:mmp}).

Case Two--$({\mathop{\rm Im}\nolimits} \left( \eta  \right) = 0$,  ${\mathop{\rm Re}\nolimits} \left( \eta  \right) = 0)$:  If the condition (\ref{eq:cccc}) is not satisfied, the equation  (\ref{eq:fsp}) has the solution with non-identical entries. Moreover, the union graph of all switching topologies is connected means that the eigenvector associated with eigenvalue zero of ${\sum\limits_{r = 1}^{\left| \mathfrak{S} \right|} {{\alpha _r}{\mathcal{L}_r}}}$ is the only ${\mathbf{1}_n}$,
for any ${\alpha _r} > 0,\sum\limits_{r = 1}^{\left| \mathfrak{S} \right|} {{\alpha _r} = 1}$. Therefore, from (\ref{eq:mmp}) with $\eta = 0$, we obtain a feasible non-zero vector $a\left( \rho  \right)$, which completes the proof of necessary condition.
\end{IEEEproof}

\begin{rem}
The strategy~(\ref{eq:cccc}) in Theorem~\ref{thm:sms} implies that the minimum number of monitored agents required to detect ZDA is equivalent to the number of components of union difference graph. Take the difference graph in Figure~\ref{fig:cpp} as an example, which has four components: two nontrivial components and two trivial components. Therefore, if the topology set $\mathfrak{S}$ includes only Graph One and Graph Two in Figure~\ref{fig:cpp}, ${\left| \mathbb{M} \right|} \geq 4$.
\end{rem}

\section{Attack Detection Algorithm}
Based on the obtained detectability of ZDA, this section focus on its detection algorithm.

\subsection{Luenberger Observer under Switching Topology}
We now present a Luenberger observer~\cite{lbo} for the system~(\ref{eq:oofn}):
\begin{subequations}
\begin{align}
\!\!\!\dot{\mathbf{x}}_{i}\!\left( t \right) &\!=\! \mathbf{v}_{i}\!\left( t \right)\label{eq:fl1}\\
\!\!\!\dot{\mathbf{v}}_{i}\!\left( t \right) &\!=\!\!  \sum\limits_{i = 1}^n\!\! {a_{ij}^{{\sigma}\left( t \right)}\!\!\left(\! {{\mathbf{x}_{j}}\!\left( t \right) \!-\! {\mathbf{x}_{i}}\!\left( t \right)} \!\right)}  \!\!-\! \!\left\{ \begin{array}{l}
\hspace{-0.22cm}\psi_{i}\mathbf{r}_{i}(t) \!+\! \theta_{i}\dot{\mathbf{r}}_{i}(t)\!, i \!\in\! \mathbb{M}\\
\hspace{-0.22cm}0, \hspace{0.0cm} i \!\in\! {\mathbb{V}} \backslash \mathbb{M}\\
\end{array} \right. \label{eq:fl2}\\
\!\!\!\mathbf{r}_{i}(t) &\!=\! \mathbf{x}_{i}(t) - \tilde{y}_{i}(t), i \!\in\! {\mathbb{V}} \backslash \mathbb{M}\label{eq:fl3}
\end{align}\label{eq:fl}\end{subequations}
where $\breve{y}_{i}(t)$ is the output of monitored agent $i$ in system~(\ref{eq:oofn}), $\mathbf{r}_{i}\left( t \right)$ is the attack-detection signal, $\psi_{i}$ and $\theta_{i}$ are the observer gains designed by the defender, $\mathbf{x}(t_{0}) = {\breve{x}}(t_{0})$, $\mathbf{v}(t_{0}) = {\breve{v}}(t_{0})$.

We define the tracking errors as ${\mathbf{e}_x}\left( t \right) \triangleq \mathbf{x}\left( t \right) - \tilde{x}\left( t \right)$ and ${\mathbf{e}_v}\left( t \right) \triangleq v\left( t \right) - \tilde{v}\left( t \right)$. A dynamics of tracking errors with attack-detection signal is obtained from~(\ref{eq:fl}) and~(\ref{eq:oofn}):
\begin{subequations}
\begin{align}
{{\dot{\mathbf{e}}}_x}\left( t \right) &= {\mathbf{e}_v}\left( t \right),\\
{{\dot{\mathbf{e}}}_v}\left( t \right) &=  - \left( {{\mathcal{L}_{\sigma \left( t \right)}} \!+\! \Phi } \right){\mathbf{e}_x}\left( t \right) \!-\! \Theta{{\mathbf{e}}_v}\left( t \right) \!-\! a\left( t \right),\\
r\left( t \right) &= C{\mathbf{e}_x}\left( t \right),
\end{align}\label{eq:fab1}\end{subequations}
where $a\left( t \right)$ is defined in~(\ref{eq:ndm4}), ${r}\left( t \right) \triangleq \mathbf{y}\left( t \right) - \tilde{y}(t)$, and 
\begin{align}
\Phi  &\triangleq \emph{\emph{diag}}\{ {\psi_{1}, \ldots, \psi_{\left| \mathbb{M} \right|}, 0, \ldots, 0} \} \in \mathbb{R}^{n \times n},\label{eq:flvb1}\\
\Theta &\triangleq \emph{\emph{diag}}\{{\theta_{1}, \ldots, \theta_{\left| \mathbb{M} \right|}, 0, \ldots, 0} \} \in \mathbb{R}^{n \times n}.\label{eq:flvb2}
\end{align}

The strategy~(\ref{eq:s2}) in Theorem~\ref{thm:sms} implies that if the union difference graph is connected, i.e., the union difference graph has only one component, using only one monitored agent's output is sufficient to detect ZDA. The following result regarding the stability of system~(\ref{eq:fab1}) in the absent of attack will answer the question: \emph{under what condition only one monitored agent's output is sufficient for the observer~(\ref{eq:fl}) to asymptotically track the system~(\ref{eq:oofn}) in the absent of attack? }
\begin{thm}
Consider the following matrix:
\begin{align}
\mathcal{A}_{s} \triangleq \left[
    \begin{array}{c;{1.0pt/1.0pt}c}
        {\mathbf{0}_{n\times n}} & {\mathbf{1}_{n\times n}}  \\ \hdashline[1pt/1pt]
        { - \mathcal{L}_{s} - \Phi} & {-\Theta}
    \end{array}
\right],\label{eq:dm}
\end{align}
where $\mathcal{L}_{s}$ is the Laplacian matrix of a connected undirected graph, the gain matrices $\Phi$ and $\Theta$ defined in~(\ref{eq:flvb1}) and~(\ref{eq:flvb2}) satisfy
\begin{align}
\mathbf{0}_{n \times n} \neq \Phi \geq 0,\label{eq:aaa}\\
\mathbf{0}_{n \times n} \neq \Theta \geq 0.\label{eq:bbb}
\end{align}
$\mathcal{A}_{s}$ is Hurwitz for any $\left| \mathbb{M} \right| \ge 1$, if and only if $\mathcal{L}_{s}$ has distinct eigenvalues. \label{thm:my0bd}
\end{thm}

\begin{IEEEproof}
See Appendix C.
\end{IEEEproof}

\subsection{Strategic Switching Topology For Detection}
The observer~(\ref{eq:fl}) can also be modeled as a switched system as well. Let us recall a technical lemma that can address the problem: \emph{when the topology of observer~(\ref{eq:fl}) should strategically switch such that it can asymptotically track the real system~(\ref{eq:oofn}) in the absence of attacks}.
\begin{lem}~\cite{porfiri2008fast} Consider the switched systems:
\begin{align}
\dot x\left( t \right) = {\mathcal{A}_{\sigma \left( t \right)}}x\left( t \right).\nonumber
\end{align}
under periodic switching, i.e., $\sigma \left( t \right)$ $=$ $\sigma \left( {t + \tau} \right) \in \mathfrak{S}$.
If there exists a convex combination of some matrix measure that satisfies
\begin{align}
\sum\limits_{m = 1}^{L} {{\nu _m}\mu\left( {{\mathcal{A}_{m}}} \right)}  < 0,\label{eq:sdsbk1}
\end{align}
then the switched system system is uniformly asymptotically stable for every positive $\tau$.
\label{thm:pss}
\end{lem}

The strategic topology-switching algorithm is described by Algorithm~2.
\begin{algorithm}
  \caption{Strategic Topology-Switching Algorithm}
  \KwIn{Initial index $k$ = 0, initial time $t_{k} = 0$, an ordered topology set $\mathfrak{S}$ that satisfies~(\ref{eq:sdsbk1}) and
  \begin{align}
  \exists s \in \mathfrak{S}: \mathcal{L}_{s} \hspace{0.1cm} \emph{\emph{has}} \hspace{0.1cm} \emph{\emph{distinct}} \hspace{0.1cm} \emph{\emph{eigenvalues}},\label{eq:bss2}
  \end{align}
  dwell times $\tau_{s}$, $s \in \mathfrak{S}$, generated by~(\ref{eq:addt}) that satisfy 
  \begin{align}
  \sum\limits_{s \in \mathfrak{S}} {{\nu _s}{\mu _P}\left( {{{\cal A}_s}} \right)}  < 0,{\nu _s} = \frac{{{\tau _s}}}{{\sum\limits_{r \in \mathfrak{S}} {{\tau _r}} }}.\label{eq:bss2}
  \end{align}}
  Run the multi-agent system~(\ref{eq:oofn}) and the observer~(\ref{eq:fl})\;
  Switch topology of system~(\ref{eq:oofn}) and its observer~(\ref{eq:fl}) at time $t_{k} + \tau_{{\sigma}(t_{k})}$: $\sigma ({t_k}) \leftarrow \mathfrak{S}\left( \!\!\!\!{\mod \!\!\!\left( {k + 1,\left| \mathfrak{S} \right|} \right) + 1} \right)$\;
  Update topology-switching time: $t_{k} \leftarrow t_{k} + \tau_{{\sigma}(t_{k})}$\;
  Update index: $k \leftarrow k+1$\;
  Go to Step 2.
\end{algorithm}

\begin{thm}
Consider the multi-agent system~(\ref{eq:oofn}) and the observer~(\ref{eq:fl}), where the observer gain matrices $\Phi$ and $\Theta$ satisfy~(\ref{eq:aaa}) and~(\ref{eq:bbb}), and the topology-switching signal ${\sigma}(t_{k})$ of the observer~(\ref{eq:fl}) and the system~(\ref{eq:oofn}) are generated by Algorithm~2.
\begin{description}
  \item[i)] Without knowledge of the misbehaving agents and the attack-starting time, the observer~(\ref{eq:fl}) is able to detect ZDA in system~(\ref{eq:oofn}), i.e., $r(t) \equiv \mathbf{0}_{\left| \mathbb{M} \right|}$ does not holds, if and only if the set of monitored agents and switching topologies satisfy~(\ref{eq:cccc}).
  \item[ii)] In the absence of attacks, without constraint on the magnitudes of observer gains, the observer~(\ref{eq:fl}) asymptotically track the real system~(\ref{eq:oofn}), i.e., the system~(\ref{eq:fab1}) is globally uniformly asymptotically stable.
  \item[iii)]In the absence of attacks, the agents in system~(\ref{eq:oofn}) achieve the asymptotic second-order consensus.
\end{description}
\label{thm:dfd}
\end{thm}
\begin{IEEEproof}
We first note that Line 3 and 5 of Algorithm~2 imply the topology-switching signal ${\sigma}(t)$  is time-dependent. 

\emph{Proof of i):} Replacing ${A}_{{{\sigma}}(t)}$ by ${\mathcal{A}}_{{{\sigma}}(t)}$ (defined in~(\ref{eq:dm})) in the steps to derive~(\ref{eq:ss1})--(\ref{eq:ss4}) in the proof of Theorem~\ref{thm:sms}, we have
\begin{align}
\eta \mathbf{e}_{x}\left( \rho \right) - \mathbf{e}_{v}\left( \rho \right) &= \mathbf{0}_{n},\label{eq:ss1aa}\\
{a}\left( \rho \right) + {\mathcal{L}_r}\mathbf{e}_{x}\left( \rho \right) + \eta \mathbf{e}_{x}(\rho) &= \mathbf{0}_{n}, \!\forall r \in \mathfrak{S}\label{eq:ss2aa}\\
{a}\left( \rho \right) + {\mathcal{L}_s}\mathbf{e}_{x}\left( \rho\right) + \eta \mathbf{e}_{x}(\rho) &= \mathbf{0}_{n}, \!\forall s \in \mathfrak{S}\label{eq:ss3aa}\\
{\mathbf{e}_{x_j}}\left( \rho \right) &= 0, \forall j \in \mathbb{M}.\label{eq:ss4aa}
\end{align}
Therefore, the rest of the proof of i) follows that of Theorem~\ref{thm:sms} straightforwardly.  

\emph{Proof of ii):} In the absence of attacks, the system matrix of system~(\ref{eq:fab1}) is $\mathcal{A}_{\sigma(t)}$ defined in~(\ref{eq:dm}). Considering~(\ref{eq:bss2}), by Lemma~\ref{thm:my0bd}, and the conditions~(\ref{eq:aaa}) and~(\ref{eq:bbb}),  ${\cal A}_{s}$ is Hurwitz. Thus, there exists $P > 0$ such that ${\mu _P}\left( \mathcal{A}_{s} \right) < 0$. Through setting on the switching times (dwell times) by~(\ref{eq:addt}), (\ref{eq:sdsbk1}) can be satisfied. By Lemma~\ref{thm:pss},  the switched linear system (\ref{eq:fab1}) is uniformly asymptotically stable, which is, in fact, equivalent to  globally uniformly asymptotically stable.

\emph{Proof of iii)} follows Lemma~\ref{thm:lmr} straightforwardly.
\end{IEEEproof}

\section{Simulation}
\begin{figure}[http]
\centering
\includegraphics[scale=0.47]{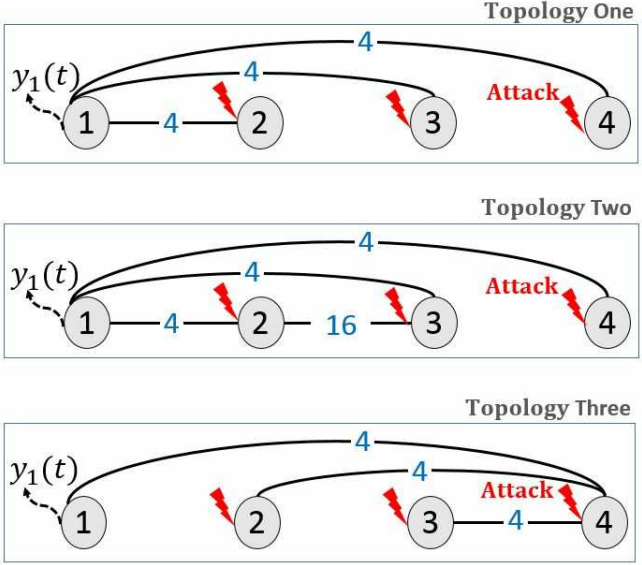}
\caption{Working situation.}
\label{fig:dp}
\end{figure}

We consider a system with $n = 4$ agents. The initial position and velocity conditions are chosen randomly as ${x}(0) = {v}(0) = {\left[ {1,2,3,4} \right]^ \top }$. The considered network topologies with their coupling weights are given in Figure~\ref{fig:dp}. The working situation is illustrated by Figure~\ref{fig:dp} as:
\begin{itemize}
  \item Agents 2--4 are misbehaving agents, i.e., $\mathbb{K} = \left\{ {{2},{3},{4}} \right\}$;
  \item only agent 1 is the monitored one, i.e., $\mathbb{M} = \left\{ {{1}} \right\}$.
\end{itemize}

Property ii) in Theorem~\ref{thm:dfd} states that our strategic topology switching has no constraint on the magnitudes of observer gains in tracking real system. To demonstrate this, we set the observer gains significantly small as
\begin{align}
\Phi = \Theta = \emph{\emph{diag}}\{10^{-6},0,0,0\}.\label{eq:ffgg}
\end{align}

\subsection{Undetectable Zero-Dynamics Attack}

First, we consider the topology set $\mathfrak{S} = \{1, 2\}$, and set the topology switching sequence as $1 \to 2 \to 1 \to 2 \to  \ldots$, periodically. It verifies from Figure~\ref{fig:dp} that the topology set $\mathfrak{S} = \{1, 2\}$ satisfies~(\ref{eq:bss1}) and~(\ref{eq:bss2}). By Lemma~\ref{thm:lmr}, we select the dwell times $\tau_{1} =  \tau_{2} = \frac{T_{1}}{2} + 0.2 = \frac{T_{2}}{2} + 0.2$ = $\frac{\pi}{2} + 0.2$.

It verifies from Topologies One and Two in Figure~\ref{fig:dp} that their generated difference graph is disconnected. Thus, the set $\mathfrak{S} = \{1, 2\}$ does not satisfy~(\ref{eq:cccc}) in Theorem~\ref{thm:sms}. Therefore, the attacker can easily design a ZDA such that the observer~(\ref{eq:fl}) under Algorithm~2 fails to detect it.

Let the attacker's goal be to make the system working under Algorithm~2 unstable, without being detected. Following the policy~(\ref{eq:ssczeo}) and the attack-starting time selection scheme--Algorithm 1, one of its attack strategies is designed as:
\begin{itemize}
  \item $\eta = 0.0161$;
  \item modify the data of initial condition sent to observer~(\ref{eq:fl}) as $\widehat{x}\left( 0 \right)$ = ${\left[ {1,1,3,5} \right]^\top}$ and $\widehat{v}\left( 0 \right)$ = ${\left[ {1,1,4,4} \right]^\top}$;
  \item choose attack-starting time $\rho = 1097.4$;
  \item introduce ZDA signal to system at $\rho$: \\
  $g\left( t \right) \!=\! {10^{ - 3}}{\left[ {0,{\rm{7}}{\rm{.3}}{e^{\eta (t - \rho )}},{\rm{7}}{\rm{.3}}{e^{\eta (t - \rho )}}, - {\rm{14}}{\rm{.6}}{e^{\eta (t - \rho )}}} \right]^\top}$.
\end{itemize}

\begin{figure}[http]
\centering{
\begin{minipage}[b]{0.63\textwidth}
\includegraphics[width=0.82\textwidth]{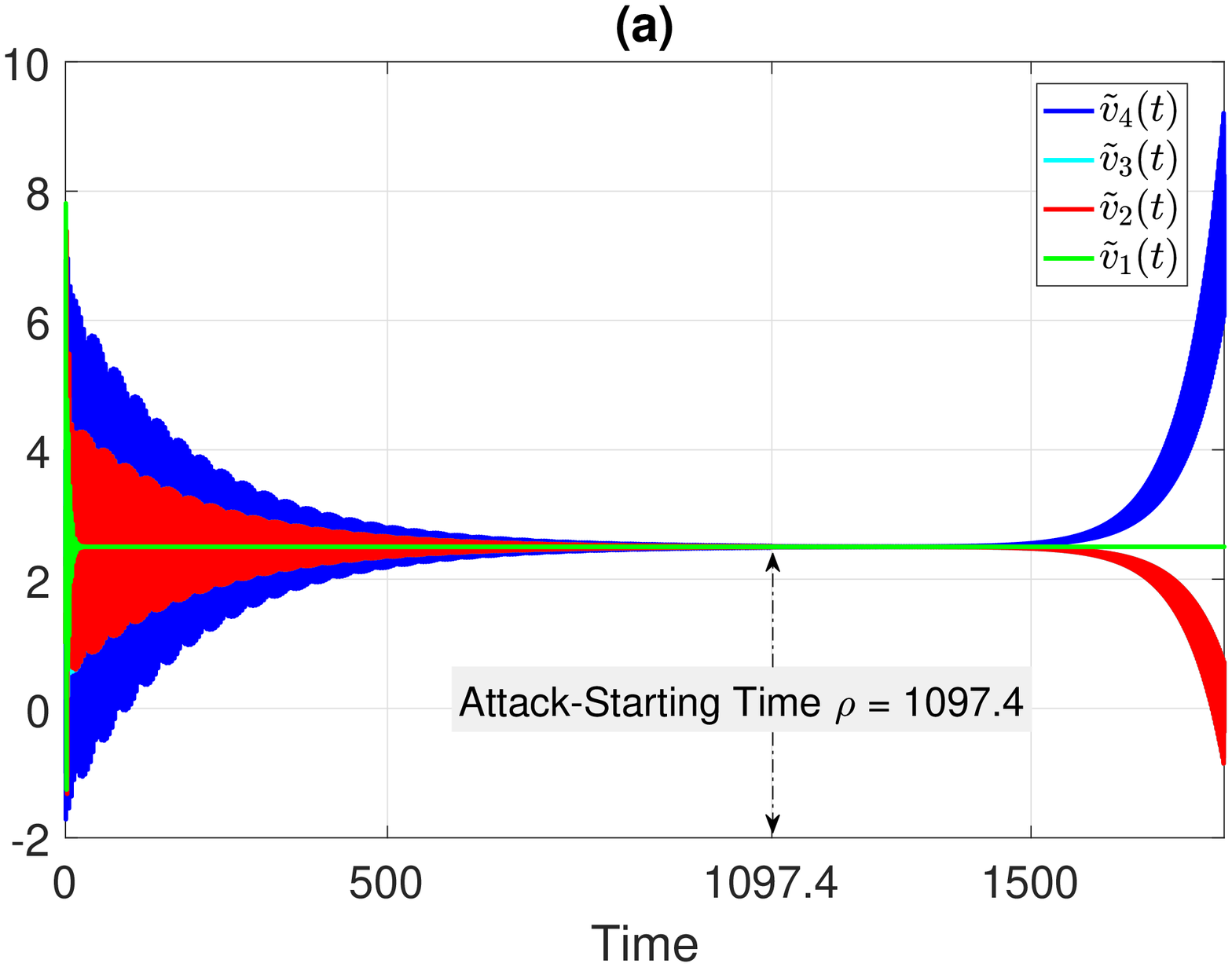} \\
\includegraphics[width=0.82\textwidth]{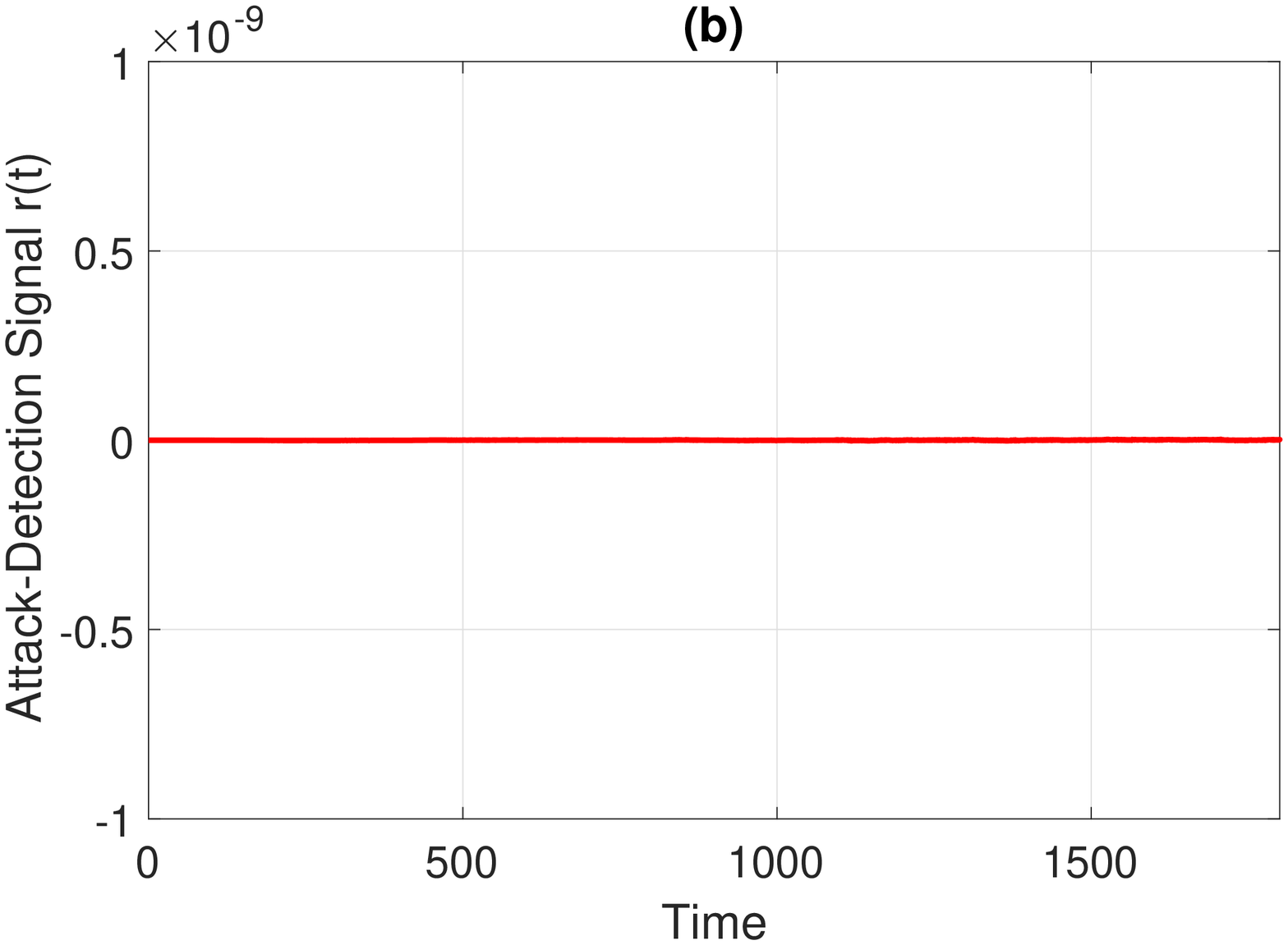}
\end{minipage}}
\caption{States $\tilde{v}(t)$: multi-agent system under attack is unstable; attack-detection signal $r(t)$: the attack keeps stealthy over time.}
\label{fig:trp}
\end{figure}

The trajectories of detection signal $r(t)$ designed in~(\ref{eq:fl}), and the velocities are shown in Figure~\ref{fig:trp}, which illustrates that the attacker's goal of making the system unstable without being detected is achieved under the topology set $\mathfrak{S} = \{1, 2\}$.

\subsection{Detectable Zero-Dynamics Attack}
To detect the designed stealthy attack, we now incorporate Topology Three in Figure~\ref{fig:dp} into topology set, i.e., $\mathfrak{S} = \{1, 2,3\}$. We let the topology switching sequence to be $1 \to 2 \to 3 \to 1 \to 2 \to 3 \to \ldots$, periodically. It verifies that the topology set $\mathfrak{S} = \{1, 2,3\}$ satisfies~(\ref{eq:bss1}) and~(\ref{eq:bss2}). Using Lemma~\ref{thm:lmr}, the dwell times are selected as $\tau_{1} =\tau_{2} = \tau_{3} = \frac{\pi}{2} + 0.2$.

We note that in the working situation illustrated by Figure~\ref{fig:dp}, the existing results~\cite{n11,n12,F13,weerakkody2017robust,chen2017protecting} for the multi-agent systems under fixed topology fail to detect ZDA. This is mainly due to the misbehaving-agents set $|\mathbb{K}|= 3$; the connectivities of Topologies One, Two and Three are as the same as $1$; and the output set $|\mathbb{M}| = 1$. All these violate the conditions on the connectivity of the communication network, the size of the misbehaving-agent set,  and the size of the output set, which are summarized in Table~I.

It verifies from Figure~\ref{fig:dp} that the difference graph generated by Topologies One and Three, or  Topologies Two and Three is connected. Thus, by property i) in Theorem~\ref{thm:dfd}, we conclude that using only one monitored agent's output, the observer~(\ref{eq:fl}) working under Algorithm~2 is able to detect the designed ZDA under the topology set $\mathfrak{S} = \{1, 2,3\}$.

\begin{figure}[http]
\centering{
\begin{minipage}[b]{0.63\textwidth}
\includegraphics[width=0.82\textwidth]{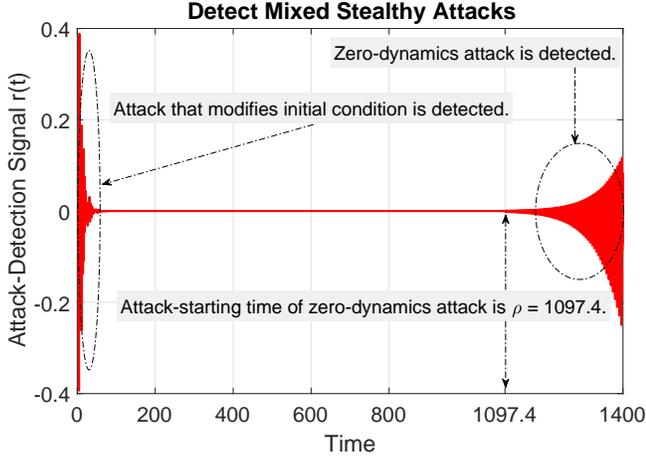}
\end{minipage}}
\caption{Attack-detection signal $r(t)$: using only one monitored agent's output, the designed ZDA is detected.}
\label{fig:rev}
\end{figure}

The trajectory of the attack-detection signal $r(t)$ is shown in Figure~\ref{fig:rev}. Remark~\ref{thm:dfd} states that when the starting time of ZDA is not the initial time, the proposed attack policy includes two mixed stealthy attacks. Figure~\ref{fig:rev} illustrates that using only agent 1's output, the mixed stealthy attacks are successfully detected.

\subsection{Observer in The Absence of Attacks}
\begin{figure}[http]
\centering{
\begin{minipage}[b]{0.63\textwidth}
\includegraphics[width=0.85\textwidth]{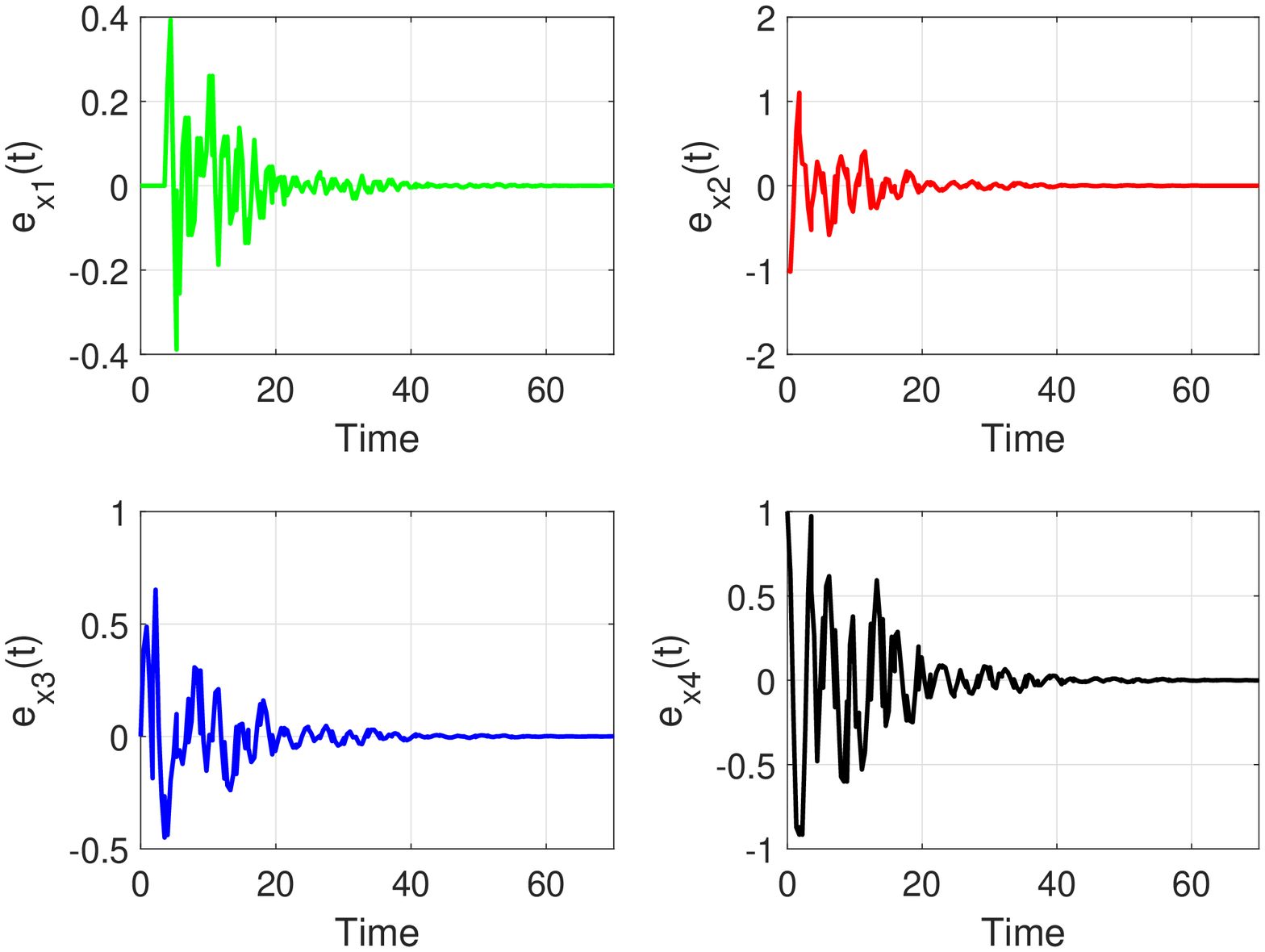} \\
\includegraphics[width=0.85\textwidth]{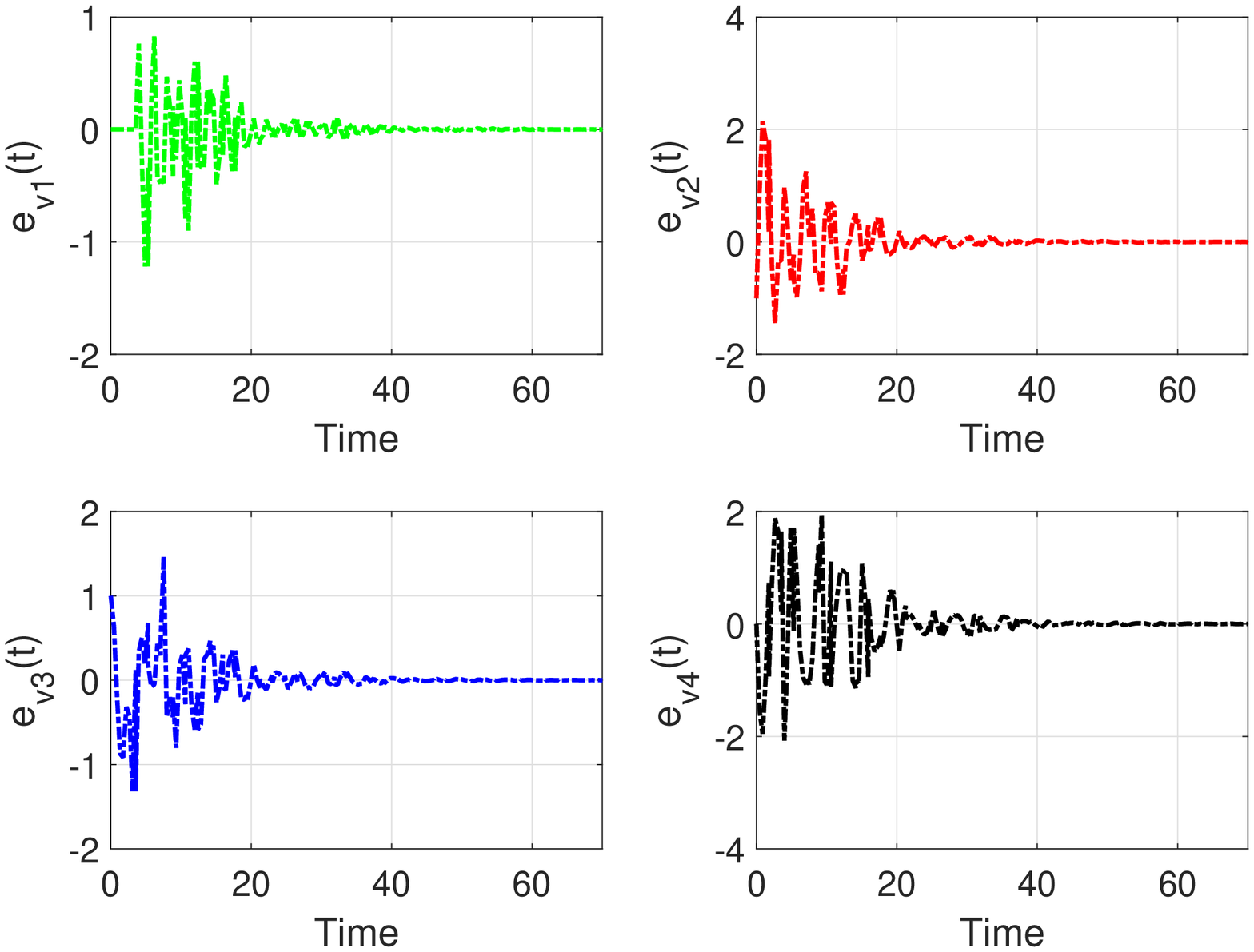}
\end{minipage}}
\caption{In the absence of attacks, trajectories of observer errors under Algorithm~2 are globally uniformly asymptotically stable.}
\label{fig:gh}
\end{figure}

We now show the effectiveness of strategic topology switching for the observer~(\ref{eq:fl}) in estimating the states of the multi-agent system~(\ref{eq:oonp}), i.e., the multi-agent system~(\ref{eq:oofn}) in the absence of attacks. Input the initial conditions modified by attacker to the observer~(\ref{eq:fl}), i.e., $\mathrm{x}\left( 0 \right)$ = $\widehat{x}\left( 0 \right)$ = ${\left[ {1,1,3,5} \right]^\top}$ and $\mathrm{v}\left( 0 \right) = \widehat{v}\left( 0 \right)$ = ${\left[ {1,1,4,4} \right]^\top}$. The trajectories of observer errors are shown in Figure~\ref{fig:gh}, which shows that using the significantly small observer gains in~(\ref{eq:ffgg}) and only agent 1's output, Algorithm~2 works successfully for the observer~(\ref{eq:fl}) to asymptotically track the real multi-agent system in the absence of attacks.
\section{Conclusion}
This two-part paper studies strategic topology switching for the second-order multi-agent system under attack.
In Part-I paper~\cite{YA}, for the simplified control protocol that does need velocity measurements, we propose a strategy on switching times that addresses the problem: \emph{when the topology should switch such that the changes in system dynamics do not undermine agent's ability of reaching the second consensus in the absence of attacks}. In Part-II paper, we propose a strategy on switching topologies that addresses the problem: \emph{what topology to switch to, such that the ZDA can be detected}. Based on the two strategies, a defense strategy is derived in this Part-II paper, its merits are summarized as
\begin{itemize}
  \item In achieving the second-order consensus in the absence of attacks, the control protocol does not need the velocity measurements, while the algorithm has no constraint on the magnitudes of coupling weights.
  \item In tracking real systems in the absence of attack, it has no constraint on the magnitudes of observer gains of the proposed Luenberger observer and the number of monitored agents.
  \item In detecting ZDA, the algorithm has no constraint on the size of misbehaving-agent set, while the algorithm allows the defender to have no knowledge of the attack-starting time. 
\end{itemize}

The theoretical results obtained in this two-part paper imply several rather interesting results:
\begin{itemize}
  \item for the size of switching topology set, there exists a fundamental tradeoff between the topology connection cost and the convergence speed to consensus;
  \item for the dwell time of switching topologies, there exist a tradeoff between the switching cost and the duration of attacks going undetected, and the convergence speed to consensus.
\end{itemize}
 Analyzing the tradeoff problems in the lights of game theory and multi-objective optimization constitutes a part of our future research.

\appendices
\section{Proof of Corollary~\ref{thm:rezda}}
From the attack signal~(\ref{eq:asg}), we know that ${g}(t) = \mathbf{0}_{\bar{o}}$ for $t \in [ {0,\rho} )$. Thus, before the attack-starting time $\rho$, the system~(\ref{eq:sl1}) is described by
\begin{subequations}
\begin{align}
\dot{\tilde z}\left( t \right) &= A\tilde z\left( t \right),\\
\tilde y\left( t \right) &= C\tilde z\left( t \right),t \in [ {0,\rho} )
\end{align}\label{eq:psl1}\end{subequations}
from which the first item in~(\ref{eq:rs2}) is obtained by integration.

It is straightforward to derive from~(\ref{eq:psl1}) and~(\ref{eq:sl2}) that
\begin{subequations}
\begin{align}
\dot{\tilde z}\left( t \right) - \dot z\left( t \right) &= A\left( {\tilde z\left( t \right) - z\left( t \right)} \right)\\
\tilde y\left( t \right) - y\left( t \right) &= C\left( {\tilde z\left( t \right) - z\left( t \right)} \right),t \in [0,\rho ).
\end{align}\label{eq:psl2}\end{subequations}

We note that~(\ref{eq:rs1}) implies ${\tilde y\left( t \right) - y\left( t \right)} = C\left( {\tilde{z}\left( t \right) - z\left( t \right)} \right) = \mathbf{0}_{\bar{m}}$ for all  $t \in [{0,\rho})$, which means $\tilde{z}(0) - z(0) \neq \mathbf{0}_{\bar{n}}$ is unobservable under the dynamics~(\ref{eq:psl2}). Thus, $\tilde{z}\left( 0 \right) - z\left( 0 \right) \in \ker \left( {{\mathcal{O}}} \right)$, i.e., the policy~(\ref{eq:aczeo}) holds.

Considering the fact that the system~(\ref{eq:sl1}) under the attack signal~(\ref{eq:asg}) is continuous w.r.t. time, which implies $\tilde{z}(\rho^{-}) = \tilde{z}(\rho)$. Therefore, $\tilde{z}(\rho) - z(\rho) = e^{A\rho} (\tilde{z}(0) - z(0))$ can be obtained from~(\ref{eq:psl2}). The results~(\ref{eq:rs1}) and~(\ref{eq:rs2}) over the time interval $[\rho, \infty)$ are generated through launching the classical ZDA signal ${g}(t) = {g}{e^{\eta \left( {t - \rho } \right)}},t \ge \rho$, the detailed proof can be found in~\cite{G93}, it is omitted here.

\section{Proof of Proposition~\ref{thm:asmsaa}}
Under the result of time-dependent topology switching~(\ref{eq:tde}),  from~(\ref{eq:s1}) and~(\ref{eq:s2}), we obtain a dynamics:
\begin{subequations}
\begin{align}
\dot{{{{\breve{z}}}}}\left( t \right) &= A_{{{\sigma}}(t)}{{{\breve{z}}}}\left( t \right), \\
\breve y\left( t \right) &= C\breve z\left( t \right), t \in [0, \rho)
\end{align}\label{eq:pos2}\end{subequations}
where $\breve{z}$ and ${\breve{y}}\left( t \right)$ are given in (\ref{eq:krs1}) and~(\ref{eq:krs2}), respectively. It follows from the dynamics~(\ref{eq:pos2}) that the solution~(\ref{eq:pal22}) is obtained by integration.

Without loss of generality, we let $[0, \rho] = \left[ {0,{t_1}} \right) \cup \left[ {{t_1},{t_2}} \right) \cup  \ldots  \cup [ {{t_k},\rho} )$ with $\rho  \le {t_{k + 1}}$, $k \in {\mathbb{N}_0}$. (\ref{eq:pos2}) implies the stealthy property: $y\left( t \right) = \tilde{y}\left( t \right)$ for all $t \in [0, \rho]$, is equivalent to
\begin{align}
{\breve{y}}\left( t \right) = \mathbf{0}_{\left| \mathbb{M} \right|}, \emph{\emph{for}} \hspace{0.1cm} \emph{\emph{all}} \hspace{0.1cm} t \in \left[ {0,{t_1}} \right) \cup  \ldots  \cup [ {{t_k},\rho } ).\label{eq:apos2}
\end{align}

We note that~(\ref{eq:apos2}) means that the system~(\ref{eq:pos2}) is unobservable for any $t \in [t_{0}, t^+_{k})$, $k \in \mathbb{N}_{0}$, which is further equivalent to $\breve{z}\left( 0 \right) \in \mathbf{N}^{k}_{1}$ with $\mathbf{N}^{k}_{1}$ recursively computed by~(\ref{eq:hhp1}) and~(\ref{eq:hhp1aa}), via considering Theorem 1 of~\cite{tanwani2013observability}. 
The condition in Lines 1 and 2 of Algorithm~1 means that once the attacker finds~(\ref{eq:apos2}) does not hold at $t_{k}$, he must immediately launches ZDA signal to keep stealthy, i.e., $\rho = t_{k}$; otherwise, ${\breve{y}}\left( t_{k} \right) \neq \mathbf{0}_{\left| M \right|}$. If~(\ref{eq:apos2}) holds,  the attacker can launch the ZDA at future time, i.e., $\rho > t_{k}$.

The set defined in~(\ref{eq:hhp2}) and the condition ``$\mathop {\max }\limits_{t \in {\mathcal{S}_{k}}} \left\{ t \right\} = {t_{k + 1}}$" contained in Line 6 of Algorithm~1 implies that the selection of $\rho$ also dependents on whether ZDA policy~(\ref{eq:czeo}) is feasible at $\rho$, so that its stealthy property can continue to hold. Line 7 of Algorithm 1 implies that if it is feasible at the incoming switching time $t_{k+1}$, the attacker can launch the ZDA signal at current activated time interval $[t_{k}, t_{k+1})$ or next interval $[t_{k+1}, t_{k+2})$. Otherwise, the attacker must launch the ZDA at a time in the current time interval, i.e., $\rho \in \mathcal{S}_{k}$.

\section{Proof of Theorem~\ref{thm:my0bd}}

We let $\sigma(t) = s \in \mathfrak{S}$ for $t \in [t_{k}, t_{k+1}), k \in \mathbb{N}_{0}$. Since ${\mathcal{L}}_s$ is the Laplacian matrix of a connected graph and $\Phi \geq 0$, ${{\mathcal{L}}_s} + \Phi$ is positive definite. We define the following positive function for the system~(\ref{eq:fab1}) with $a(t) \equiv \mathbf{0}_{n}$:
\begin{align}
{V_s}\!\left( {e\left( t \right)} \right) \!=\! \frac{1}{2}e_x^\top\!\left( t \right)\!\left( {{\mathcal{L}}_s} \!+\! \Phi \right){e_x}\!\left( t \right) \!+\! e_v^\top\!\left( t \right){e_v}\!\left( t \right).\nonumber
\end{align}
Its time derivative is obtained as
\begin{align}
{{\dot V}_s}\left( {e\left( t \right)} \right) =  - e_v^\top\left( t \right) \Theta e_v\left( t \right) \le 0,\label{eq:hs1}
\end{align}
where the inequality is obtained by considering $\Theta \geq 0$. Since the dynamics~(\ref{eq:fab1}) with $a(t) \equiv \mathbf{0}_{n}$ is equivalent to $\dot e\left( t \right) = {{\cal A}_s}e\left( t \right)$ with $e\left( t \right) \triangleq \left[\!\!
    \begin{array}{c;{1pt/1pt}c}
        e_x^\top\left( t \right) & e_v^\top\left( t \right)
    \end{array}
\!\!\right]^\top$, we conclude from~(\ref{eq:hs1}) that none of the eigenvalues of $\mathcal{A}_{r}$ has positive real part.

We next prove $\mathcal{A}_{r}$ has neither zero nor pure imagine eigenvalues.

Using the well-known formula $\det  \left( \left[
    \begin{array}{c;{1.0pt/1.0pt}c}
        A & B  \\ \hdashline[1pt/1pt]
        C & D
    \end{array}
\right] \right)$  $=$ $\det \left( A \right)\det \left( {D - C{A^{ - 1}}B} \right)$, from~(\ref{eq:dm}) we have:
\begin{align}
&\det \left( {\mathcal{A}_{s} - \lambda {\mathbf{1}_{2n \times 2n}}} \right) \nonumber\\
&= \det \left( \left[
    \begin{array}{c;{1.0pt/1.0pt}c}
        {-\lambda\mathbf{1}_{n\times n}} & {\mathbf{1}_{\left| \mathbb{V} \right|\times \left| \mathbb{V} \right|}}  \\ \hdashline[1pt/1pt]
        {- \mathcal{L}_{s} - \Phi} & - \Theta - \lambda{\mathbf{1}_{n \times n}}
    \end{array}
\right]\right)\nonumber\\
& = \det \left( { - \lambda {\mathbf{1}_{n \times n}}} \!\right)\det \left({ -\Theta - \lambda {\mathbf{1}_{n \times n}} - \frac{ \mathcal{L}_{s} + \Phi}{\lambda}} \right)\nonumber\\
& = \det \left( {{\lambda^2}{\mathbf{1}_{n \times n}} + \Theta\lambda + \mathcal{L}_{s} + \Phi} \right). \label{eq:pdm1}
\end{align}

Let us define:
\begin{align}
{\phi _m} &\triangleq \sqrt {\psi_{m} + \lambda \theta_{m}} {{\mathbf{e}}_m},\label{eq:pa1}
\end{align}
with ${\mathbf{e}}_m \in {\mathbb{R}^{n}}$ being the $m^{\emph{\emph{th}}}$ vector of the canonical basis. It verifies from~(\ref{eq:pa1}),~(\ref{eq:flvb1}) and~(\ref{eq:flvb2}) that
\begin{align}
 \mathcal{P}(m) \triangleq \lambda\widehat{\Theta}(m)  + \widehat{\Phi}(m) = \sum\limits_{p = m}^{\left| \mathbb{M} \right|} {{\phi _p}} \phi _p^\top ,\label{eq:paa2}
\end{align}
with
\begin{align}
\widehat{\Theta}(m) &\triangleq \emph{\emph{diag}}\left\{ {0, \ldots ,0,\theta_{m}, \ldots ,\theta_{\left| \mathbb{M} \right|},0, \ldots 0} \right\}, \nonumber\\
\widehat{\Phi}(m) &\triangleq \emph{\emph{diag}}\left\{ {0, \ldots ,0,\psi_{m}, \ldots ,\psi_{\left| \mathbb{M} \right|},0, \ldots 0} \right\}. \nonumber
\end{align}

Let us recall the well-known formula:
\begin{align}
\det \left( {A + \chi u{w^\top}} \right) = \det \left( A \right)\left( {1 + \chi{w^\top}{A^{ - 1}}u} \right), \label{eq:paa3}
\end{align}
where $A$ is invertible, and $w$ and $u$ are vectors. By~(\ref{eq:paa3}), we obtain from~(\ref{eq:pdm1}) and~(\ref{eq:pa1})--(\ref{eq:paa3}) that
\begin{align}
&\det \left( {{\mathcal{A}_s} - \lambda {\mathbf{1}_{2n \times 2n}}} \right) \nonumber\\
& = \prod\limits_{m = 1}^{\left| \mathbb{M} \right|} {\left( {1 + \phi _m^ \top {{\left( {{\mathcal{H}_s} + \mathcal{P}\left( {m + 1} \right)} \right)}^{ - 1}}{\phi _m}} \right)} \det \left( {{\mathcal{H}_s}} \right),\label{eq:pdm3}
\end{align}
where
\begin{align}
{\mathcal{H}_s} \triangleq {\lambda ^2}{{\bf{1}}_{_{n \times n}}} + {{\cal L}_s}.\label{eq:pdm3a}
\end{align}

Since $\mathcal{L}_r$ is a symmetric matrix, there exists an orthogonal matrix ${Q}$ $\triangleq$ $\left[ {{q_1}; \ldots; {q_n}} \right]$ $\in$ ${\mathbb{R}^{\left| \mathbb{V} \right| \times \left| \mathbb{V} \right|}}$ with ${q_i}$ $\triangleq$ $\left[
    \begin{array}{c;{1pt/1pt}c;{1pt/1pt}c;{1pt/1pt}c}
        {q_{i1}} & {q_{i2}} & \ldots & q_{i\left| \mathbb{V} \right|}
    \end{array}
\right]^\top$ $\in$ $\mathbb{R}^{\left| \mathbb{V} \right|}$, $i \in \mathbb{V}$, such that
\begin{subequations}
\begin{align}
{Q^ \top } &= {Q^{ - 1}},\label{eq:wp1}\\
{Q^\top}\!{\mathcal{H}_s}Q &= \emph{\emph{diag}}\left\{ {{\lambda ^2} + {\lambda _1}({{\cal L}_s}), \ldots ,{\lambda ^2} + {\lambda _{n}}({{\mathcal{L}}_s})} \right\}.\label{eq:wp3}
\end{align}\label{eq:wpxx}
\end{subequations}
Considering~(\ref{eq:bbb}) and~(\ref{eq:flvb2}), without loss of generality, we let
\begin{align}
\theta_{\left| \mathbb{M} \right|} \neq 0.\label{eq:new1}
\end{align}

It follows from~(\ref{eq:wpxx}) and~(\ref{eq:pa1}) that
\begin{align}
\phi _{\left| \mathbb{M} \right|}^ \top {\left( {{\mathcal{H}_s}} \right)^{ - 1}}{\phi _{\left| \mathbb{M} \right|}} &= \sum\limits_{i = 1}^{n} {\frac{{\left( {\psi_{\left| \mathbb{M} \right|} + \lambda \theta _{\left| \mathbb{M} \right|}} \right)q_{i\left| \mathbb{M} \right|}^2}}{{{\lambda _i}\left( {{\mathcal{L}_s}} \right) + {\lambda ^2}}}},\label{eq:pdm4} \\
\det \left( {{\mathcal{H}_s}} \right) &= \prod\limits_{i = 1}^{n} {\left( {{\lambda _i}\left( {{\mathcal{L}_s}} \right) + {\lambda ^2}} \right)},\label{eq:pdm4a}
\end{align}
from which, we arrive at
\begin{align}
&\left( {1 + \phi _{\left| \mathbb{M} \right|}^ \top {{\left( {{\mathcal{H}_s}} \right)}^{ - 1}}{\phi _{\left| \mathbb{M} \right|}}} \right)\det \left( {{\mathcal{H}_s}} \right) \nonumber\\
& = \prod\limits_{i = 1}^{n} {\left( {{\lambda _i}\left( {{\mathcal{L}_s}} \right) + {\lambda ^2}} \right)} \nonumber\\
& \hspace{0.4cm} + \sum\limits_{i = 1}^{n} {\prod\limits_{j \ne i}^{n} {\left( {{\lambda _j}\left( {{\mathcal{L}_s}} \right) + {\lambda ^2}} \right)} \left( {\psi_{\left| \mathbb{M } \right|} + \lambda \theta_{\left| \mathbb{M} \right|}} \right)q_{i\left| \mathbb{M} \right|}^2}. \label{eq:pdm5}
\end{align}

Let us define:
 \begin{align}
\mathcal{Q}\left( \lambda  \right) \triangleq \prod\limits_{m = 1}^{\left| \mathbb{M} \right| - 1} {\left( {1 + \phi _m^ \top {{\left( {{{\cal H}_s} + {\cal P}\left( {m + 1} \right)} \right)}^{ - 1}}{\phi _m}} \right)}.\label{eq:pdm5a}
\end{align}
Substituting (\ref{eq:pdm5}) and~(\ref{eq:pdm5a}) into~(\ref{eq:pdm3}) yields
\begin{align}
&\det \left( {{\mathcal{A}_s} - \lambda {{\bf{1}}_{2n \times 2n}}} \right) \nonumber\\
& = \mathcal{Q}\left( \lambda  \right)\!\!\left( {\prod\limits_{i = 1}^{n} {\left( {{\lambda _i}\left( {{{\cal L}_s}} \right) + {\lambda ^2}} \right)} } \right. \nonumber\\
&\hspace{0.4cm}\left. { + \sum\limits_{i = 1}^{n} {\prod\limits_{j \ne i}^{n} {\left( {{\lambda _j}\left( {{{\cal L}_s}} \right) + {\lambda ^2}} \right)} \!\left( {\psi_{\left| \mathbb{M} \right|}  \!+\! \lambda \theta_{\left| \mathbb{M} \right|}} \right)q_{i\left| \mathbb{M} \right|}^2} } \right).\label{eq:aap1}
\end{align}
In the followings, we consider two different cases.

\subsection{Case One: $\mathcal{A}_{r}$ has zero eigenvalue}
In this case, i.e., $\lambda = 0$, it follows from~(\ref{eq:pa1}),~(\ref{eq:paa2}),~(\ref{eq:pdm3a}),~(\ref{eq:pdm5a}) and the condition $\theta_{m} \geq 0$ and $\psi_{m} \geq 0$, $\forall m \in \mathbb{M}$, that $\mathcal{Q}\left( \lambda  \right) > 0$. Thus, we conclude from~(\ref{eq:aap1}) that ${\left. {\det \left( {{{\cal A}_s} - \lambda {{\bf{1}}_{2n \times 2n}}} \right)} \right|_{\lambda  = 0}} > 0$. Therefore, $\mathcal{A}_{r}$ does not have any zero eigenvalue. A contradiction occurs.

\subsection{Case Two: $\mathcal{A}_{r}$ has pure imagine eigenvalue}
This case means $\lambda = \varpi\mathrm{i}$ with $0 \neq \varpi \in \mathbb{R}$. It verifies from~(\ref{eq:pa1}),~(\ref{eq:paa2}) and~(\ref{eq:pdm3a})  that
\begin{align}
1 + \phi _m^ \top {\left( {{\mathcal{H}_s} + P\left( {m + 1} \right)} \right)^{ - 1}}{\phi _m} \ne 0, \forall m \in \mathbb{M} \nonumber
\end{align}
thus, ${\left. {\mathcal{Q}\left( \lambda  \right)} \right|_{\lambda  = \varpi \mathrm{i}}} \ne 0$. Then, we conclude from~(\ref{eq:aap1}) that $\det \left( {{{\cal A}_s} - \mathrm{i}\varpi {{\bf{1}}_{2n \times 2n}}} \right) = 0$ is equivalent to
\begin{align}
&\prod\limits_{i = 1}^{n} \!{\left( {{\lambda _i}\!\left( {{\mathcal{L}_s}} \right) \!-\! {\varpi ^2}} \right)}  \!+\! \sum\limits_{i = 1}^{n} {\prod\limits_{j \ne i}^{n}\! {\left( {{\lambda _j}\!\left( {{\mathcal{L}_s}} \right) \!-\! {\varpi ^2}} \right)\psi_{\left| \mathbb{M} \right|}q_{i\left| \mathbb{M} \right|}^2} } \nonumber\\
&\hspace{0.4cm} + \mathrm{i}\sum\limits_{i = 1}^{n} {\prod\limits_{j \ne i}^{n} {\left( {{\lambda _j}\left( {{\mathcal{L}_s}} \right) - {\varpi ^2}} \right)} } \varpi \theta_{\left| \mathbb{M} \right|}q_{i\left| \mathbb{M} \right|}^2 = 0.\label{eq:pdm6xx}
\end{align}

We note that~(\ref{eq:pdm6xx}) implies
\begin{align}
\sum\limits_{i = 1}^{n} {\prod\limits_{j \ne i}^{n} {\left( {{\lambda _j}\left( {{\mathcal{L}_s}} \right) - {\varpi ^2}} \right)} } \varpi \theta_{\left| \mathbb{M} \right|}q_{i\left| \mathbb{M} \right|}^2 = 0,
\end{align}
which, in conjunction with~(\ref{eq:new1}), results in
\begin{align}
\sum\limits_{i = 1}^{n} {\prod\limits_{j \ne i}^{n} {\left( {{\lambda _j}\left( {{\mathcal{L}_s}} \right) - {\varpi ^2}} \right)} } \varpi q_{i\left| \mathbb{M} \right|}^2 = 0, \nonumber
\end{align}
which further implies that
\begin{align}
\sum\limits_{i = 1}^{n} {\prod\limits_{j \ne i}^{n}\! {\left( {{\lambda _j}\!\left( {{\mathcal{L}_s}} \right)  \!-\! {\varpi ^2}} \right)\psi_{\left| \mathbb{M} \right|}q_{i\left| \mathbb{M} \right|}^2} } = 0.\nonumber
\end{align}
Thus, from~(\ref{eq:pdm6xx}) we have $\prod\limits_{i = 1}^{n} \!{\left( {{\lambda _i}\!\left( {{\mathcal{L}_s}} \right) - {\varpi ^2}} \right)} = 0$, which means that
\begin{align}
&\exists i \in \{1,\ldots,n\}\!: \hspace{0.2cm}\varpi^{2} = {{\lambda _i}\left( {{{\cal L}_s}} \right)}.\label{eq:bss2xx}
\end{align}

However, it is straightforward to verify from~(\ref{eq:bss2xx}) that $\sum\limits_{i = 1}^{n} {\prod\limits_{j \ne i}^{n} {\left( {{\lambda _j}\left( {{\mathcal{L}_s}} \right)  - {\varpi ^2}} \right)} } \varpi \theta_{\left| \mathbb{M} \right|}q_{i\left| \mathbb{M} \right|}^2  \neq 0$ if and only if $\mathcal{L}_{s}$ has distinct eigenvalues. Consequently, (\ref{eq:pdm6xx}) does hold, thus a contradiction occurs.

\section*{Acknowledgment}
The authors thank Dr. Sadegh Bolouki, Dr. Hamidreza Jafarnejadsani, and Dr. Pan Zhao for valuable discussions.

\ifCLASSOPTIONcaptionsoff
  \newpage
\fi

\bibliographystyle{IEEEtran}
\bibliography{refII}
\end{document}